\documentclass[useAMS,usenatbib]{mn2e}

\usepackage{txfonts}
\usepackage{graphicx} %
\usepackage{natbib} %
\usepackage{array,tabularx}

\IfFileExists{srcltx.sty}{\usepackage[active]{srcltx}}

\usepackage{bm,paralist,xspace}

\newcommand{\eprint}[1]{\href{http://arxiv.org/abs/#1}{#1}}

\newcommand{\adsurl}[1]{\href{#1}{ADS}}

\usepackage[breaklinks=true]{hyperref} %

\IfFileExists{hypernat.sty}{\usepackage{hypernat}}

\newcommand{\GeV}{\ensuremath{\:\mathrm{GeV}}} %
\newcommand{\MeV}{\ensuremath{\:\mathrm{MeV}}} %
\newcommand{\keV}{\ensuremath{\:\mathrm{keV}}} %
\newcommand{\kev}{\ensuremath{\:\mathrm{keV}}} %
\newcommand{\eV} {\ensuremath{\:\mathrm{eV}}}   %
\newcommand{\cm}{\:\mathrm{cm}} 
\newcommand{\pc}{\:\mathrm{pc}} 
\newcommand{\kpc}{\:\mathrm{kpc}} 
\newcommand{\mpc}{\:\mathrm{Mpc}} 
\newcommand{\dm}{{\textsc{dm}}} 
\newcommand{\mdm}{m_{s}} 
\newcommand{\xmm}{\textsl{XMM-Newton}\xspace}

\newcommand{\eq}[1]{\begin{equation} #1 \end{equation}}

\begin{document}

\title[Constraints on decaying Dark Matter from M31]{Constraints on decaying
  Dark Matter from \textsl{XMM-Newton}
  observations of M31} %

\author[A.~Boyarsky et al.]{Alexey~Boyarsky$^{a,b}$, Dmytro~Iakubovskyi$^b$,
  Oleg~Ruchayskiy$^c$, and\ Vladimir~Savchenko$^{b,d}$\\
  \it $^{a}$PH-TH, CERN, CH-1211 Geneve 23,
  Switzerland\\
  $^b$Bogolyubov Institute for Theoretical Physics, Kiev
  03780, Ukraine\\
  $^{c}$\'Ecole Polytechnique F\'ed\'erale de Lausanne, Institute of
  Theoretical Physics, FSB/ITP/LPPC, BSP 720, CH-1015, Lausanne,
  Switzerland\\
  $^{d}$Physics Department, Kiev National Taras Shevchenko University, Kiev
  03022, Ukraine }

\date{Received $<$date$>$ ; in original form $<$date$>$ }

\pagerange{\pageref{firstpage}--\pageref{lastpage}} %
\pubyear{2007}

\maketitle

\label{firstpage}

\begin{abstract}
  We derive constraints on the parameters of the radiatively decaying Dark Matter
  (DM) particle, using the \textsl{XMM-Newton} EPIC spectra of the Andromeda
  galaxy (M31). Using the observations of the outer (5'-13') parts of M31, we
  improve the existing constraints. For the case of sterile neutrino DM,
  combining our constraints with the latest computation of abundances of
  sterile neutrinos in the Dodelson-Widrow (DW) scenario, we obtain the lower
  mass limit $m_{s} < 4 \keV$, which is stronger than the previous one $m_s <
  6 \kev$, obtained recently by~\citet{Asaka:06c}. Comparing this limit with
  the most recent results on Lyman-$\alpha$ forest analysis of~\citet{Viel:07}
  ($m_s > 5.6\kev$), we argue that the scenario in which all the DM is
  produced via the DW mechanism is ruled out.  We discuss however other production
  mechanisms and note that the sterile neutrino remains a viable candidate
  for Dark Matter, either warm or cold.
\end{abstract}

\section{Introduction}

A vast body of evidence points to the existence of Dark Matter
(DM) in addition to the ordinary visible matter in the Universe.
The evidence includes: velocity curves of galaxies in clusters and
stars in galaxies; observations of galaxy clusters in X-rays;
gravitational lensing data; cosmic microwave background
anisotropies. While the DM contributes some 22\% to the total
energy density in the Universe, its properties remain largely
unknown.

The Standard Model of particle physics (SM) does not provide a DM
candidate.  The DM cannot be made out of baryons, as such an
amount of baryonic matter cannot be generated in the framework of
an otherwise successful scenario of Big Bang
nucleosynthesis~\citep{Dar:95}. In addition, current microlensing
experiments exclude the possibility that MACHOs (massive compact
halo objects) constitute the dominant amount of the total mass
density in the local halo~\citep{Gates:95,EROS:00,MACHO:00}.  The
only possible non-baryonic DM candidate in the SM could be the
neutrino, however this possibility is ruled out by the present
data on the large scale structure (LSS) of the Universe.

What properties of the DM particles can be deduced from existing
observations? Some information comes from studies of structure
formation. Namely, the velocity distribution of the DM particles
at the time of structure formation affects greatly the power
spectrum of density perturbations, as measured by a variety of
experiments (see e.g.~\citealt{Tegmark:03}).  One of the
parameters, characterizing the influence of the DM velocity
dispersion on the power spectrum, is the \emph{free-streaming
length} $\lambda_{FS}$ -- the distance traveled by the DM particle
from the time when it became non-relativistic until today. Roughly
speaking, the free-streaming length determines the minimal scale
at which the Jeans instability can develop, and therefore
non-trivial free-streaming implies modification of the spectrum of
density perturbations at wave numbers $k\gtrsim
\lambda_{FS}^{-1}$.

If the DM particles have negligible velocity dispersion, they
constitute the so-called \textit{cold} DM (CDM), which forms
structure in a ``bottom-up'' fashion (i.e. smaller scale objects
formed first and then merged into the larger ones, see
e.g.~\citealt{Peebles:80}). The neutrino DM represents the
opposite case -- \emph{hot} DM (HDM). In HDM scenarios, structure
forms in a top-down fashion~\citep{Zeldovich:70}, and the first
structures to collapse have size comparable to the Hubble
scale~\citep{Bisnovatyi:80,Bond:80,Doroshkevich:81,Bond:83}. In
this scenario the galaxies do not have enough time to form,
contradicting to the existing observations \citep[see
e.g.][]{White:83,Peebles:84a}.

\emph{Warm} DM (WDM) represents an intermediate case, cutting
structure formation at some scale, with the details being
dependent on a particular WDM model.

Both WDM and CDM fit the LSS data equally well. The differences
appear when one starts to analyze the details of structure
formation for galaxy-size objects (modifications of the power
spectrum at momenta $k\gtrsim 0.5 h \mpc^{-1}$). It is usually
said that WDM predicts ``less power at smaller scales'', meaning
in particular that one expects smaller number of dwarf satellite
galaxies and shallower density profiles than those predicted by
CDM models~\citep{Navarro:96,Klypin:99,Ghigna:99}.  Thus WDM
models can provide the way to solve the ``missing satellite''
problem and the problem of central density peaks in galaxy-sized
DM halos \citep{Klypin:99,Moore:99,Bode:00,Avila-Reese:01}.

There exist a number of direct astrophysical observations which
seem to contradict the N-body simulations of galaxy formations,
performed in the framework of the CDM
models~\citep[e.g.][]{Diemand:06,Strigari:07}.  Namely, direct
measurements of the DM density profiles in dwarf spheroidal (dSph)
satellites of the Milky Way favour cored
profiles~\citep{Gilmore:06,Gilmore:07a,Wu:07,Gilmore:07b}.\footnote{For
  certain dSph cusped profiles are still admissible, but disfavored.
  Additional considerations rule out the possibility of existence of cusped
  profiles for the Ursa Minor and
  Fornax~\citep{Kleyna:03,Kleyna:03b,Goerdt:06,Sanchez:06b}.}
The number of dwarf satellite galaxies, as currently observed, is
still more than an order of magnitude below the CDM predictions,
in spite of the drastically improved sensitivity towards the
search \citep[see][]{Gilmore:07a,Koposov:07} and resolution of
numerical simulations~\citep{Strigari:07}. There seems to exist a
smallest scale (${\sim} 120$~pc) at which the DM is
observed~(\citealt{Gilmore:07a,Gilmore:07b}). However, as of now
there is no definitive statement about the ``CDM substructure
crisis'' (see \cite{Simon:07} in regard to the smallest observed
DM scale and~\cite{Penarrubia:07} for an alternative solution of
the ``missing satellite problem'').

The power-spectrum of density perturbations at scales of interest
for the WDM vs. CDM issue can also be studied, analyzing the
Lyman-$\alpha$ forest data (absorption feature by the neutral
hydrogen at $\lambda = 1216$\AA\ at different red-shifts in the
distant quasar spectra,~\citealt{Hui:97}). This involves
comparison of the observed spectra of Ly-$\alpha$ absorption lines
with those obtained as a result of numerical simulations in
various DM models. In this way one arrives at an upper limit on
the free-streaming length of the DM particles.

Various particle physics models provide WDM candidates.  Possible
examples include gravitinos and axinos in various supersymmetric
models (see e.g.~\citealt{Baltz:01,Cembranos:06,Seto:07}). Another
WDM candidate is the sterile neutrino with a mass in the keV
range~\citep{Dodelson:93}. Recently, this candidate received a lot
of attention.  Namely, an extension of the minimal SM (MSM) with
the three right-handed neutrinos was
suggested~\citep{Asaka:05b,Asaka:05a}. This extension (called
$\nu$MSM) explains several \textit{observed} phenomena beyond the
MSM under the minimal number of assumptions. Namely, apart from
the absence of the DM candidate, the MSM fails to explain observed
\textit{neutrino oscillations} -- the transition between neutrinos
of different flavors~(for a review see
e.g.~\citealt{Fogli:05,Strumia:06,Giunti:06}). The explanation of
this phenomenon is the existence of neutrino mass.  The most
natural way to provide this mass is to add right-handed neutrinos.
Indeed, in the MSM, neutrinos are left-handed (all other fermions
have both left-handed and right-handed counterparts) and strictly
massless.  The structure of the MSM dictates that right-handed
neutrinos, if added to the theory, would not be charged with
respect to any Standard Model interactions and interact with other
matter only via mixing with the usual (left-handed) neutrinos
(that is why right-handed neutrinos are often called
\textit{sterile neutrinos} to distinguish them from the
left-handed \textit{active} ones).  Moreover, as demonstrated by
\citealt{Asaka:05b}, the parameters of added right-handed
neutrinos can be chosen in such a way that such a model resolves
another problem of the MSM -- it explains the excess of baryons
over antibaryons in the Universe (the \textit{baryon asymmetry}),
while at the same time it does not spoil the predictions of Big
Bang nucleosynthesis. For this to be true, the masses of two of
these sterile neutrinos should be chosen in the range $300 \MeV
\lesssim M_{2,3} \lesssim 20 \GeV$, while the mass of the third
(lighter) sterile neutrino is arbitrary (as long as it is below
$M_{2,3}$). In particular, its mass can be in the keV range,
providing the WDM candidate. Such a sterile neutrino can be
produced in the Early Universe in the correct amount via various
mechanisms: via non-resonant oscillations with active
neutrinos~\citep{Dodelson:93,Dolgov:00,Abazajian:01a,Asaka:06b,Asaka:06c},
via interaction with the inflaton \citep{Shaposhnikov:06}, via
resonant oscillations in the presence of lepton
asymmetries~\citep{Shi:98}, and have cosmologically long
life-time.

Finally, the sterile neutrino with mass in the \keV range would
have other interesting astrophysical applications (see e.g.
\citet{Sommer:99,Kusenko:06a,Biermann:06,Hidaka:06,Hidaka:07,Stasielak:06}
and references therein).

\subsubsection*{Existing bounds on sterile neutrino DM}
\label{sec:existing-bounds}

The mass of the sterile neutrino DM should satisfy the universal
Tremaine-Gunn lower bound \citep{Tremaine:79,Dalcanton:00}: $\mdm
\ge 300-500\eV$. A stronger (although model dependent) lower bound
comes from the Lyman-$\alpha$ forest analysis. Assuming a
particular velocity distribution of the sterile
neutrino\footnote{Sterile neutrinos are not in thermal equilibrium
in the early Universe and therefore their velocity distribution is
non-universal and depends on the model of production.} %
one can obtain a relation between the DM mass and $\lambda_{FS}$
and therefore convert an upper bound on the free-streaming length
to a \emph{lower} bound on the mass of the sterile neutrino.  In
the recent works of \citet{Seljak:06,Viel:06} this bound was found
to be 14 keV (correspondingly 10 keV) at 95\% CL in the
Dodelson-Widrow (DW) production model~\citep{Dodelson:93}. New
results from QSO lensing give similar restrictions for the DW
model: $\mdm \ge 10$\keV~\citep{Miranda:07}.  For different models
of production, the relation between the DM mass and the
free-streaming length is different and the Lyman-$\alpha$ mass
bound for sterile neutrinos can be as low as $M_s >2.5$ keV~(see
e.g.~\citealt{Ruchayskiy:07}).\footnote{Strictly speaking, in case
of other models of production the power spectrum of density
fluctuations is not only characterized by the free-streaming
length. Therefore, the rescaling of the results
of~\citet{Seljak:06,Viel:06} can be used only as the estimates and
the reanalysis of the Lyman-$\alpha$ data for the case of each
model is required.}

The sterile neutrino DM is not completely stable. In particular,
it has a radiative decay channel into an active neutrino and a
photon, emitting a monoenergetic photon with energy $E_\gamma =
m_s/2$ (where $m_s$ is the mass of the sterile neutrino). As a
result, the (indirect) search for the DM decay line in the X-ray
spectra of objects with large DM overdensity becomes an important
way to restrict the parameters (mass and decay width) of sterile
neutrino DM. During the last two years a number of papers appeared
devoted to this task:
\citealt{Boyarsky:05,Boyarsky:06b,Boyarsky:06c,Boyarsky:06e,Boyarsky:06f,Riemer:06,Watson:06,Boyarsky:06d,Abazajian:06b}.
The current status of these observations is summarized, e.g.,
in~\cite{Ruchayskiy:07}. The results of the computation of sterile
neutrino production in the early Universe~\citep{Asaka:06c},
combined with these X-ray bounds, puts an upper bound on the
sterile neutrino mass of $m_s < 6\kev$~\citep{Asaka:06c}. This is
below the \emph{lower} bound on the sterile neutrino DM mass from
the Lyman-$\alpha$ forest analysis of~\citet{Seljak:06,Viel:06}.
Thus it would seem that the scenario, in which all the sterile
neutrino DM is produced via the DW mechanism, is ruled out (the
recent work by~\cite{Palazzo:07} also explored the possibility
that the sterile neutrino, produced through DW scenario,
constitutes but a fraction of DM and found this fraction to be
below 70\%). However, the results of~\citet{Seljak:06,Viel:06} are
based on the low-resolution SDSS Lyman-$\alpha$ dataset
of~\citet{McDonald:05}. It was shown recently by~\cite{Viel:07}
that using high-resolution HIRES spectra~\citep{Becker:06} one
arrives at the lower limit $m_s > 5.6\kev$.  Thus, the small
window of masses $5.6\kev < m_s < 6 \kev$ remains open in the DW
model. Therefore further improvement of X-ray bounds is crucial
for exploring (and possibly closing) this region of parameters.

It was shown in~\citet{Boyarsky:06c} that the objects in the Local
Halo (e.g. dwarf spheroidal galaxies) are the best objects in
terms of the signal to noise ratio. The Andromeda galaxy (M31) is
one of the nearest galaxies, excluding dwarves, that enables one
to resolve most of its bright point sources and extract the
spectrum of its diffuse emission. It also has a massive and
well-studied dark matter
halo~\citep[e.g.][]{Klypin:02,Widrow:05,Geehan:06,Tempel:07}. The
first step in such an analysis was done by \citet{Watson:06}
(hereafter denoted by \textbf{W06}), who analyzed the diffuse
emission from the $5$ central arcmin, using the data processed by
\citet{Shirey:01}.  We repeat the analysis of the central part of
the M31, processing more observations, and extend the analysis to
the off-centre region ($5'-13'$). We also analyze the
uncertainties in the DM distribution in the central part of M31.
The outer region of M31 has much fainter diffuse emission than its
central part \citep[c.f. e.g.][Fig.~8]{Takahashi:04}, and
uncertainties in the determining of the distribution of DM in this
region are lower. All this allows us to strengthen the
restrictions on the parameters of sterile neutrino DM, while using
more conservative estimates of the DM signal.

The paper is organized as follows.  We briefly summarize the
properties of decaying DM in
Section~\ref{sec:decaying-dark-matter}. The description of DM in
M31 and expected DM decay flux is computed in
Section~\ref{sec:m31}. In Section~\ref{sec:reduction} we describe
the methodology of EPIC MOS and PN data reduction which we perform
by using two different methods: \emph{Extended
  Sources Analysis Software} (ESAS) and single background subtraction method
(SBS).  In Section~\ref{sec:fitting} we fit the spectra and obtain the
restrictions on sterile neutrino parameters. Finally, we discuss our results
in Section~\ref{results}.

\section{Decaying Dark Matter model}
\label{sec:decaying-dark-matter}

The flux of the DM decay from a given direction (in $\mathrm{photons\; s^{-1}
  cm^{-2}}$) is given by
\begin{equation}
  F_{DM}=\frac{\Gamma E_{\gamma}}{m_s}\int_{fov\ cone}
  \frac{\rho_{DM}(\mathbf{r})}{4\pi |\mathbf{D}_{L} +
    \mathbf{r}|^{2}}\mathbf{dr}.
\label{flux_general}
\end{equation}
Here $ \textbf{D}_{L}$ is the \textit{luminosity} distance between
an observer and the centre of an observed object,
$\rho_{DM}(\textbf{r})$ is the DM density, and the integration is
performed over the DM distribution inside the (truncated) cone --
solid angle, spanned by the field of view (FoV) of the X-ray
satellite.  In case of distant objects\footnote{Namely, if
luminosity
  distance $D_{L}$ is much greater than the characteristic scale of the DM
  distribution.}, Eq.~(\ref{flux_general}) can be simplified:
\eq{F_{DM}=\frac{M_{DM}^{fov}\Gamma}{4\pi
    D_{L}^{2}}\frac{E_{\gamma}}{m_s},\label{flux_distant}} where
$M^{fov}_{DM}$ is the mass of DM within a telescope field of view, $m_s$ --
mass of the sterile neutrino DM. In the case of small FoV,
Eq.~(\ref{flux_distant}) simplifies to \eq{F_{DM}=\frac{\Gamma S_{DM}\Omega
    E_{\gamma}}{4\pi m_{s}},} where \eq{S_{DM} = \int_{l.o.s.}
  \rho_{DM}(r)dr\label{S_DM}} is the DM column density (the integral goes
along the line of sight), $\Omega \ll 1$ - FoV solid angle.

The decay rate of the sterile neutrino DM is equal to
\citep{Pal:81,Barger:95}\footnote{Our decay rate is 2 times
smaller than the one used in W06. This is due to the Majorana
nature of the sterile
  neutrino, which we consider \citep[c.f.][]{Barger:95}. The final constraints
  for a Dirac particle would thus be 2 times stronger.}
\eq{\Gamma = \frac{9 \alpha G_{F}^{2}}{1024\pi^{4}} \sin^{2}( 2\theta)
  m_{s}^{5} \approx 1.38 \cdot 10^{-30}
  s^{-1}\left[\frac{\sin^{2}(2\theta)}{10^{-8}}\right]\left[\frac{m_{s}} {1
      \keV}\right]^{5}.} Here $m_{s}$ is the sterile neutrino mass, $\theta$ -
mixing angle between  sterile and active neutrinos. From a compact
cloud of sterile neutrino DM we therefore obtain the flux:
\eq{F_{DM}\approx 6.38
  \cdot 10^{6}\; \frac{\keV}{ \cm^2\cdot s}
  \left[\frac{M_\dm^{fov}}{10^{10}M_{\odot}}\right]\left[\frac{\kpc}{D_{L}}\right]^{2}\sin^{2}(2\theta)\left[\frac{m_{s}}{1
      \keV}\right]^{5}.\label{flux}}

\section{Andromeda galaxy (M31)}
\label{sec:m31}

M31, or Andromeda galaxy, is one of the nearest galaxies, excluding dwarves;
it is located at the distance $D_L = 784\pm 13 \pm 17\kpc$ \citep{Stanek:98}.
Its proximity allows us to resolve most of its point sources and extract the
spectrum of diffuse emission of its central part.

Available \xmm \citep{Jansen:01} observations cover the region of
central $15'$ of M31 with exposure time greater than $100$ ksec
(see Table \ref{M31_centre}). W06 used the \xmm data on central
$5'$ of M31 (observation \texttt{0112570401} processed by
\citet{Shirey:01}, exposure time about 30 ksec) to produce
restrictions on the parameters of sterile neutrino DM.  The
sufficient increase of photon statistics enables us to analyze the
outer ($5'$-$13'$) faint part of M31, which, however, has a
significant mass of DM (see Section~\ref{sec:DM_calc} below).

In this work we will analyze two different spatial regions of
Andromeda galaxy: region \texttt{circle5}, which corresponds to
$5'$ circle around the centre of M31, and region
\texttt{ring5-13}, which corresponds to the ring with inner and
outer radii of $5' $ and $13'$, respectively.

\begin{table*}
{
\begin{tabular}{|c|c|c|c|}
\hline Obs. ID & Starting time, UTC & Filter & Cleaned
MOS1/MOS2/PN exposure, ks \\ \hline
0112570401 & 2000-06-25 08:12:41 & Medium & 30.8/31.0/27.6\\
\hline
0109270101 & 2001-06-29 06:15:17 & Medium & 40.1/41.9/47.4\\
\hline
0112570101 & 2002-01-06 18:00:56 & Thin & 63.0/63.0/55.3\\
\hline
\end{tabular}
} \caption{Observations of the central part of M31, used in our
analysis.} \label{M31_centre}
\end{table*}

\subsection{Calculation of DM mass}\label{sec:DM_calc}

To obtain the restriction on parameters of the decaying DM, we
should calculate the total DM mass $M_\dm^{fov}$, which
corresponds to both spatial regions: \texttt{circle5} and
\texttt{ring5-13}, both with and without resolved point sources.
To estimate the systematic uncertainties of the evaluation of the
DM decay signal and to find the most conservative estimate for it,
we analyze various available DM profiles
\citep{Kerins:01,Klypin:02,Widrow:05,Geehan:06,Carignan:06,Tempel:07}:


\begin{table*}
{
\begin{tabular}{|l|c|c|c|c|}
\hline Model&\texttt{circle5}&\texttt{ring5-13}&13 arcmin sphere, MC result & 13 arcmin sphere, analytical result\\
\hline
K1, with sources &$3.27 \pm 0.01$ &$12.49 \pm  0.03$ &$5.84 \pm 0.02$&5.84\\
\hline
K2, with sources &$11.88 \pm 0.03$ &$23.75 \pm 0.09$ &$20.76 \pm 0.09$&-\\
\hline
GFBG, with sources &$6.59 \pm 0.02$ &$20.46 \pm 0.06$ &$13.40 \pm 0.03$&13.39\\
\hline
KING, with sources&$6.68 \pm 0.01$&$24.61 \pm 0.05$&$14.80 \pm 0.02$&14.80\\
\hline
MOORE, with sources &$7.34 \pm 0.02$ &$19.48 \pm 0.02$ &$13.79 \pm 0.02$&13.78\\
\hline
N04, with sources &$7.68 \pm 0.03$ &$22.89 \pm 0.07$ &$15.16 \pm 0.06$&15.18\\
\hline
NFW, with sources &$11.08 \pm 0.04$ &$40.5 \pm 0.1$ &$22.3 \pm 0.1$&22.25\\
\hline
BURK, with sources &$6.71 \pm 0.02$ &$27.97 \pm 0.03$ &$15.90 \pm 0.05$&15.90\\
\hline
KER, with sources &$5.35 \pm 0.02$ &$22.45 \pm 0.04$ &$11.56 \pm 0.03$&11.56\\
\hline
M31A, with sources &$5.95 \pm 0.01$ &$16.45 \pm 0.02$ &$11.03 \pm 0.02$&-\\
\hline
M31B, with sources &$4.99 \pm 0.01$ &$14.24 \pm 0.01$ &$9.40 \pm 0.02$&-\\
\hline
M31C, with sources &$5.60 \pm 0.01$ &$16.12 \pm 0.01$ &$10.29 \pm 0.02$&-\\
\hline
\end{tabular}} \caption{ DM mass (in 10$^{9}$ M$_{\odot}$) inside regions, used
in our analysis: results of our Monte Carlo integration. The point
sources are not excluded here. The 95\% statistical errors are
also shown. The DM distributions of~\citet{Klypin:02} (before and
after adiabatic contraction), \citet{Geehan:06} and
\citet{Kerins:01}
are marked as ``K1'', ``K2'', ``GFBG'' and ``KER'', respectively.
The DM distributions from \citet{Tempel:07} are marked as
``KING'', ``MOORE'', ``N04'', ``NFW'' and ``BURK'' (see text). The
DM distributions from \citet{Widrow:05} are marked as ``M31A'',
``M31B'' and ``M31C''.} \label{DM_res1}
\end{table*}


\begin{table*}
{
\begin{tabular}{|l|c|c|c|c|}
\hline Model&\texttt{circle5}&Removed from \texttt{circle5,} \%&\texttt{ring5-13}& Removed from \texttt{ring5-13}, \%\\
\hline
K1, without sources &$0.767 \pm 0.004$& 76.6 & $9.71 \pm 0.02$&22.3\\
\hline
K2, without sources &$2.31 \pm 0.02$& 80.4 &$18.09 \pm 0.08$&23.9\\
\hline
GFBG, without sources &$1.48 \pm 0.01$& 77.4 &$15.77 \pm 0.06$&23.0\\
\hline
KING, without sources&$1.64 \pm 0.01$&75.5&$18.99 \pm 0.06$&22.9\\
\hline
MOORE, without sources &$1.52 \pm 0.01$&79.2&$14.98 \pm 0.03$&23.1\\
\hline
N04, without sources &$1.70 \pm 0.02$ & 77.7 &$17.62 \pm 0.05$&23.0\\
\hline
NFW, without sources &$2.59 \pm 0.01$ & 76.7 &$31.34 \pm 0.07$&22.5\\
\hline
BURK, without sources &$1.67 \pm 0.02$&75.1 &$21.68 \pm 0.02$&22.5\\
\hline
KER, without sources &$1.33 \pm 0.01$& 75.0 & $17.42 \pm 0.04$ &22.5\\
\hline
M31A, without sources &$1.24 \pm 0.01$& 79.3 & $12.66 \pm 0.02$ &22.9\\
\hline
M31B, without sources &$1.04 \pm 0.01$& 79.1 & $10.98 \pm 0.01$ &23.0\\
\hline
M31C, without sources &$1.21 \pm 0.01$& 78.4& $12.43 \pm 0.01$& 22.9\\
\hline
\end{tabular}
} \caption{ DM mass (in 10$^{9}$ M$_{\odot}$) without point
sources: results of our Monte Carlo integration. {The fraction of
DM, removed together with the point sources, is also shown.} All
notations are the same as in previous table.} \label{DM_res2}
\end{table*}

\begin{itemize}
\item {\bf (K1)} \textit{Before}\footnote{In contrast to the other models,
    this model does not describe the current DM distribution, but helps
    our understanding the time evolution of DM mass inside constant FoV.} adiabatic
  contraction stage, \citet{Klypin:02} assume that DM distribution is purely
  Navarro-Frenk-White (NFW) \citep{Navarro:96}: \eq{\rho_{DM}(r) =
    \frac{1}{4\pi \left[\log(1+C) -
        C/(1+C)\right]}\frac{M_{vir}}{r(r+r_{s})^{2}}.\label{NFW}} The
  parameters of this NFW distribution (in terms of the favored C1 model of
  \citealt{Klypin:02}) are: $M_{vir} = 1.60 \times 10^{12}~M_{\odot}$; $r_{s}$
  = $25.0$~kpc; $C = 12$.

\item {\bf (K2)} This non-analytical model is the result of adiabatic
  contraction of the {\bf K1} profile, described above. To obtain it, we extract
  the data from the Fig.~4 of \citet{Klypin:02}.  In the top part of this
  figure the dot-dashed curve is the contribution of the DM halo to the total
  M31 mass distribution (C1 model of \citealt{Klypin:02}).  As the precise
  form of this mass distribution is not analytic, we scanned this curve and
  produced the file with numerical values of enclosed mass $M_{DM}(r)$ within
  the \textit{sphere} of radius $r$. After that, we interpolated the
  $M_{DM}(r)$, and evaluated the radial density distribution \eq{\rho_{DM}(r)
    = \frac{1}{4\pi r^{2}}\frac{dM_{DM}(r)}{dr}.}

\item {\bf (GFBG)} Preferred Navarro-Frenk-White distribution from
  \citet{Geehan:06}: $M_{vir} = 6.80 \times 10^{11}~M_{\odot}$; $r_{s} =
  8.18$~kpc; $C = 22$.

\item {\bf (KER)} Isothermal profile used in~\citet{Kerins:01}:

\eq{ \rho_{KER}(r)=\left\{
\begin{array}{ll}
\rho_{h}(0) \frac{a^{2}}{a^{2}+r^{2}}\;\;\;\; r \le R_{max},

\nonumber\\

0\;\;\;\; r > R_{max}.

\end{array}
\right. } where $\rho_{h}(0) = 0.23 M_{\odot} \pc^{-3}$, $a = 2
\kpc$, {\bf $R_{max} = 200 \kpc$}.

\item {\bf (M31A-C)} Profiles of~\citet{Widrow:05}. In this paper the authors
  propose several models, which differ by the relative disk/halo contribution.
  These non-analytical models (M31a-d) incorporate an exponential disk, a
  Hernquist model bulge, an NFW halo (before contraction) and a central
  supermassive black hole.  The stability against the formation of bars was
  numerically studied.\footnote{We do not use the fourth model (M31d), because
    in~\citet{Widrow:05} it was found that this model develops a bar, which
    rules it out experimentally.}


\end{itemize}
We also use density distributions from the recent paper of
\citet{Tempel:07}. The main aim of this paper is to derive the DM
density distribution in the central part of M31 (0.02-35~kpc from
the centre).

\begin{itemize}

\item {\bf (KING)} Modified isothermal profile~\citep{King:62,Einasto:74}:

\eq{ \rho_{ISO}(r)=\left\{
\begin{array}{ll}
\rho_{0}\left(\left[1+\frac{r^{2}}{r_{c}^{2}}\right]^{-1} -
\left[1+\frac{r_{0}^{2}}{r_{c}^{2}}\right]^{-1}\right)\;\;\;\; r
\le r_{0},

\nonumber\\

0\;\;\;\; r > r_{0}.

\end{array}
\right. } where $\rho_{0} = 0.413 M_{\odot} \pc^{-3}$, $r_{c} = 1.47 \kpc$,
{\bf $r_{0} = 117 \kpc$}.

\item {\bf (MOORE)} Moore profile~\citep{Moore:99}:

  \eq{ \rho_{MOORE}(r) =
    \frac{\rho_{c}}{\left(\frac{r}{r_{c}}\right)^{1.5}\left[1+\left(\frac{r}{r_{c}}\right)^{1.5}\right]},}
  where $\rho_{c} = 4.43 \cdot 10^{-3} M_{\odot} \pc^{-3}$, $r_{c} = 17.9
  \kpc$.

\item {\bf (N04)} Density distribution of~\citet{Navarro:04}:

  \eq{ \rho_{N04}(r) =
    \rho_{c}\exp\left[-\frac{2}{\alpha}\left(\frac{r^{\alpha}}{r_{c}^{\alpha}}
        -1 \right)\right],} where parameter $\alpha$, according to
  simulations, equals to $0.172\pm0.032$ \citep{Navarro:04}. For \textbf{N04}
  we take $\alpha = 0.17$, $\rho_{c} = 6.42 \cdot 10^{-3} M_{\odot} pc^{-3}$,
  $r_{c} = 11.6 \kpc$.

\item {\bf (NFW)} Navarro-Frenk-White profile: \eq{\rho_{NFW}(r) =
    \frac{\rho_{c}}{\frac{r}{r_{c}}\left[1+\left(\frac{r}{r_{c}}\right)^{2}\right]},}
  where $\rho_{c} = 5.20 \cdot 10^{-2} M_{\odot} \pc^{-3}$, $r_{c} = 8.31
  \kpc$.

\item {\bf (BURK)} Burkert profile~\citep{Burkert:95}:

  \eq{ \rho_{BURK}(r) =
    \frac{\rho_{0}}{\left(1+\frac{r}{r_{c}}\right)\left(1+\frac{r^{2}}{r_{c}^{2}}\right)},}
  where $\rho_{0} = 0.335 M_{\odot} \pc^{-3}$, $r_{c} = 3.43 \kpc$.

\end{itemize}

\begin{figure}
  \includegraphics[width=0.95\columnwidth,angle=0]{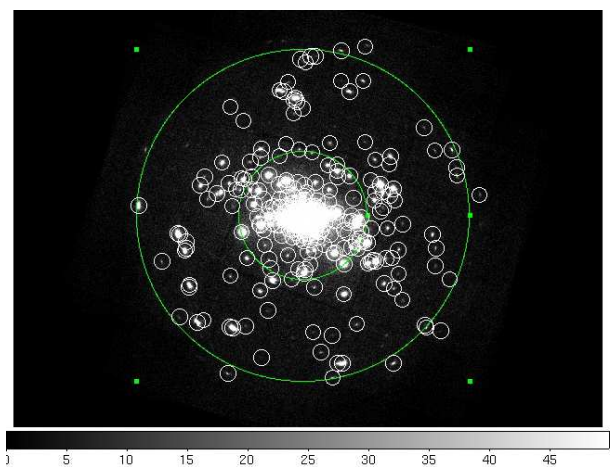}
\caption{\small Selected regions in the central part of M31 (shown
  in linear scale). Small circles correspond to excluded point source regions,
  large circles have radius of 5 and 13 arcmin.}
\label{point_sources}
\end{figure}

\begin{figure}
\includegraphics[width=0.95\columnwidth,angle=0]{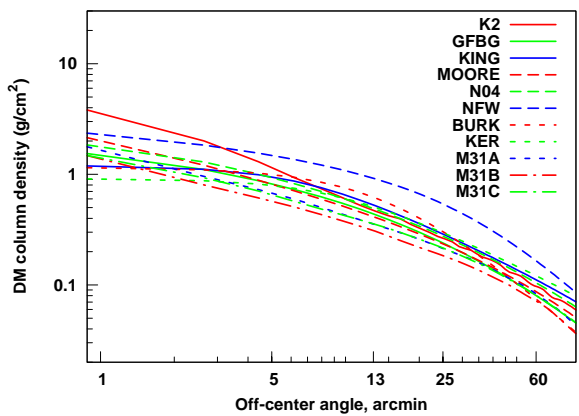}
\caption{\small M31 DM column density versus off-centre angle as
result of our Monte Carlo integration, based on DM profiles of
Sec.~\ref{sec:DM_calc}. (Point sources are not excluded).}
\label{DM_signal}
\end{figure}

The computed DM masses within the FoV for all these profiles are
shown in Table \ref{DM_res1}. We see that for the model used by
W06 (model \textbf{K2} in our notations), our estimate of the DM
mass within the central $5'$ coincides with the value used in W06:
$M_{5} = (1.3 \pm 0.2) \cdot 10^{10} M_{\odot}$.  Notice, however,
that to obtain the diffuse spectrum, we extracted all point
sources, resolved with the significance $\ge4\sigma$. Each source
was removed with the circle of the radius of $36''$ (see
Sec.~\ref{analysis_esas} for details). This led to the reduction
of the area of the FoV by about 70\% in case of \texttt{circle5}
region (c.f. Fig.~\ref{point_sources}). As the density of the DM
changes with the off-centre distance and this change can be
significant (c.f. Fig.~\ref{DM_signal}), we performed the
integration of the DM density distribution over the FoV with
excluded point sources. To calculate the DM mass in such ``swiss
cheese'' regions (Fig.~\ref{point_sources}), we used Monte Carlo
integration. The results are summarized in the
Table~\ref{DM_res2}.

To check possible systematic effects of our Monte Carlo
integration method, we also obtained the values of enclosed mass
inside the {13 arcmin} \textit{sphere}, and compared them with
analytical calculations (wherever possible). Such an error {does}
not exceed the purely statistical error of numerical integration
(see Table \ref{DM_res1}).

As one can see from Tables~\ref{DM_res1}-\ref{DM_res2}, the most
conservative DM model, describing regions \texttt{circle5} and
\texttt{ring5-13}, is the model \textbf{M31B} of
\citet{Widrow:05}.  Therefore, to obtain restrictions on the DM
parameters in what follows, we will use the DM mass estimates
based on this model.

For the DM distributions listed above, we also build the DM column
density $S_{\dm}$ {(given by Eq.~(\ref{S_DM}))} versus off-centre
angle. The result is shown on Fig.~\ref{DM_signal}. It is clearly
seen that, in the off-centre regions, there is still a lot of DM,
and, together with the fact that the surface brightness of X-ray
diffuse emission falls rapidly outside the central
$5'$~\citep[c.f.][]{Takahashi:04}, improving the restrictions of
W06 by analyzing the off-centre $5'{-}13'$ ring.  Moreover, as one
can see from Table~\ref{DM_res2} and Fig.~\ref{DM_signal}, the
uncertainty of DM in this region is less than in the
\texttt{circle5} region.

To estimate the additional contribution from the Milky Way DM halo
in the direction of M31, we use an isothermal DM distribution (as
e.g.  in \citealt{Boyarsky:06c,Boyarsky:06d}).  The DM column
density is equal to \eq{S_{MW,DM}=\frac{v_{h}^{2}}{8\pi r_c
G_{N}}K(\phi),} where $v_{h}$ = 170 km s$^{-1}$, $r_{c}$ = 4 kpc
-- parameters of isothermal model, $r_{\odot}$ = 8 kpc -- distance
from Earth to the Galactic Centre, and
\begin{equation}
K(\phi)=\frac{r_c}{R(\phi)}\left\{
\begin{array}{ll}
  \frac{\pi}{2}+\arctan\left[\frac{r_{\odot}\cos\phi}{R(\phi)}\right],
  &\cos\phi
  \ge 0 \\\\
  \arctan\left[\frac{R(\phi)}{r_{\odot}\cos\phi}\right], & \cos\phi < 0.
\end{array}
\right. \label{eq:1}
\end{equation}
Here $\phi$ is defined via $ \cos\phi = \cos l \cos b$ for an object with
galactic coordinates $(b,l)$, $R(\phi) = \bigl(r_{c}^{2}+r^2_\odot
\sin^{2}\phi\bigr)^{1/2}$.  For Andromeda galaxy ($\textit{l}=121.17^\circ$,
$\textit{b} = -21.57^\circ$, i.e. $\phi = 118.77^\circ$) one obtains
\begin{equation}
  \label{eq:2}
  S_{MW,DM} \approx 6.2 \cdot 10^{-3} g \cdot
  \cm^{-2} = 3.5 \times 10^{27}\kev\cdot\cm^{-2}
\end{equation}
According to Fig.~\ref{DM_signal}, the MW contributes $< 5\%$ to
the total DM column density along the central part of Andromeda
galaxy, and therefore will be neglected in what follows.

\section{Data reduction and background subtraction}

\label{sec:reduction}

To obtain restrictions on the parameters of the sterile neutrino,
we need to analyze diffuse emission from faint extended regions of
M31. There exist several well-developed background subtraction
procedures for the diffuse sources (see, for instance, \xmm SAS
User
Guide\footnote{\url{http://xmm.esac.esa.int/external/xmm_user_support/documentation/sas_usg/USG}},
\citealt{Nevalainen:05}, \citealt{Read:03}).  In this paper we use
two methods of background subtraction:

\subsection{Extended Sources Analysis Software
(ESAS)}\label{analysis_esas}

This method, recently developed by ESAC/GSFC team\footnote{We use ESAS version
  1.0.}, allows one to subtract instrumental and cosmic backgrounds separately. It
seems to be better than the subtraction of the scaled blank-sky
background, averaged through the entire \textsl{XMM-Newton} Field
of View (see next subsection for details), as instrumental and
cosmic backgrounds (due to their different origin) have different
vignetting correction factors. ESAS models instrumental background
from ``first principles'', using filter-wheel closed data and data
from the unexposed corners of archived observations. Using this
software, we are assured that no DM line can be in our background,
in contrast with the ``black sky'' background subtraction method
and, especially, local background subtraction (used e.g. in
\citet{Shirey:01} to produce the diffuse spectrum of central $5'$
of M31). The price to pay is the necessity of modelling cosmic
background.

To prepare the EPIC MOS \citep{Turner:01} event lists, we used the
ESAS script \texttt{mos-filter}. After running
\texttt{mos-filter}, we produced cleaned MOS images in sky
coordinates, which were used to obtain the mosaic image (with the
help of SAS v.7.0.0 tool \texttt{emosaic}). We used these event
lists and images to find the point sources using SAS task
\texttt{edetect\_chain}. Source detections were accepted with
likelihood values above 10 (about 4$\sigma$). We found 243 point
sources in this way. After that, we excluded each of them within
the circular region of the radius $36''$, which corresponds to the
removal of $\sim 70-85 \%$ of total encircled energy, depending on
the on-axis angle (see XMM users
handbook\footnote{\url{http://xmm.esac.esa.int/external/xmm_user_support/documentation/uhb}}
for details).  The constructed mosaic image with detected point
sources and selected regions is shown in Fig.~\ref{point_sources}.

We obtained the MOS1 and MOS2 spectra and constructed the
corresponding background with the help of ESAS scripts
\texttt{mos-spectra}\footnote{To
  produce correct RMF file, we changed in the script \texttt{mos-spectra}
  option \texttt{rmfgen detmaptype=psf} to \texttt{rmfgen
    detmaptype=dataset}.} and \texttt{xmm-back}, respectively.

Finally, we grouped the spectra with corresponding response and background
files with the help of FTOOL \texttt{grppha}, a part of HEASOFT v6.1. To
ensure Gaussian statistics, the minimum number of counts per bin was set to be
50.

\begin{figure}
  \includegraphics[width=0.8\columnwidth,angle=270]{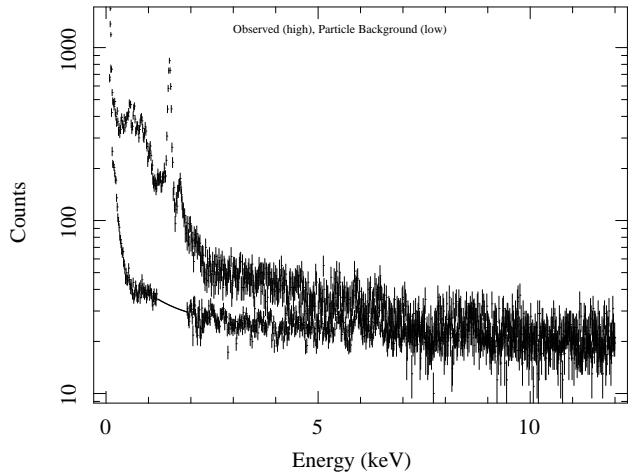}
\caption{\small Observed spectrum (top) and modelled instrumental
  background (bottom) MOS1 from ObsID 0112570101, region \texttt{ring5-13}. It
  can be seen that the spectrum and modelled background almost coincide for $E
  > 7 \keV$.} \label{sp_back}
\end{figure}

\begin{figure}
\includegraphics[width=0.75\columnwidth,angle=270]{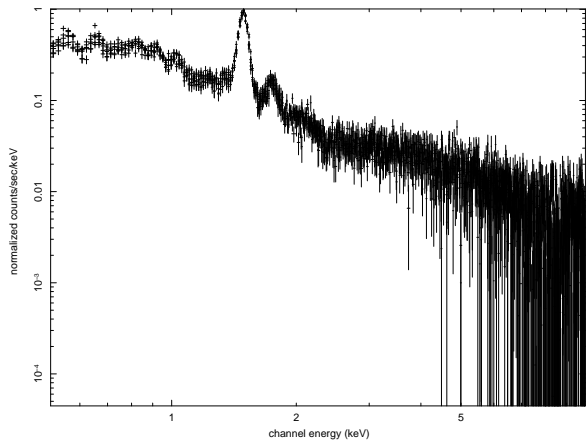}
\caption{\small Folded spectra from \texttt{ring5-13} region (by
ESAS method), with excluded point sources. The presence of two
unsubtracted instrumental lines at 1.49 \keV and 1.75 \keV is
clearly shown.} \label{r513}
\end{figure}

The ESAS method of background subtraction, however, has several
difficulties. The number of fitting parameters substantially
increases, hence it is harder to find true minimum of $\chi^{2}$.
The quantitative analysis of the $1.3{-}1.8$~keV energy range is
also not possible, because of the presence of two strong
unmodelled instrumental lines (see
Figs.~\ref{sp_back},~\ref{r513}). EPIC-PN \citep{Strueder:01} data
reduction is not yet implemented in ESAS. Therefore, to
cross-check the results obtained with the help of ESAS software,
we also processed EPIC data with the help of the blank-sky data
subtraction (SBS) method \citep{Read:03}.

\subsection{Blank-sky background subtraction (SBS)\label{SBS_proc}}

We processed the same M31 observations (Table \ref{M31_centre}) as
in the previous Section, using both MOS and PN data. To subtract
the blank-sky background we firstly cast it at the position of M31
with the help of the script \texttt{skycast}
\footnote{\url{http://www.sr.bham.ac.uk/xmm3/skycast}}, written by
the \xmm group in Birmingham. The scaling coefficient was derived
by comparing count rates for $E\ge10\keV$ from source regions and
background sample. To produce spectra, ARF, RMF and to group them
correctly (we needed to extract them from non-circular regions),
we modified the Birmingham script
\texttt{createspectra}.\footnote{\url{http://www.sr.bham.ac.uk/xmm3/createspectra}}
The spatial regions were chosen similarly to those in
Sec.~\ref{analysis_esas}, so it would be possible to compare the
results of the two different methods (see Sec.\ref{restrictions}).

When analyzing PN data, we found that the role of out-of-time
(OOT) events was significant. This is due to the fact that the
rate of OOT events is proportional to the total rate inside the
full PN FoV rather than the rate of diffuse emission (outside
excluded point sources). Therefore, it was necessary to remove the
OOT events from the PN event lists. Most of the OOT events (from
the bright point sources) form strips in the images and can be
easily removed with the help of spatial filtering. This additional
filtering also slightly reduced the possible DM signal, which was
(in this outer region) nearly proportional to \texttt{BACKSCALE}
keyword. This was accounted for when producing SBS PN
restrictions.

\section{Fitting the spectra in \texttt{Xspec} and producing restrictions}%
\label{sec:fitting}

\begin{figure}
\includegraphics[width=0.75\columnwidth,angle=270]{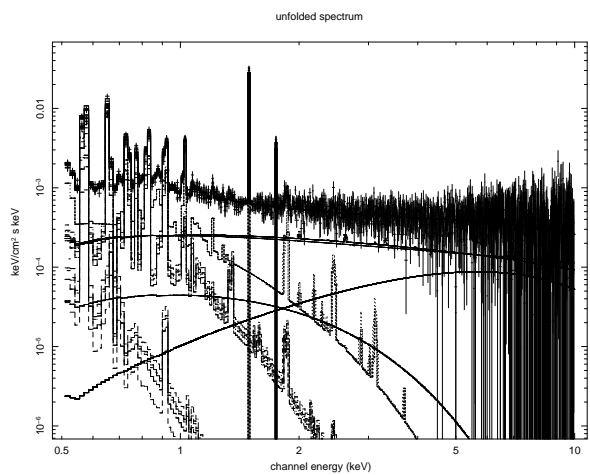}
\caption{\small Unfolded spectra and best-fit model from
\texttt{ring5-13} region (by ESAS method), with excluded point
sources. The ``line forest'' at energies lower 2.0 keV is clearly
visible.} \label{s_r513}
\end{figure}

After we have prepared the data (with ESAS and SBS background
subtraction methods) we fitted obtained spectra with realistic
model (using \texttt{Xspec} spectral fitting package version
11.3.2, \citealt{Arnaud:96}). The results of our fits are shown in
Tables~\ref{pars_c5},~\ref{chi},~\ref{abundances}. Notice that the
fit results obtained by two background subtraction methods (ESAS
and SBS) coincide within the 90\% confidence interval
(Table~\ref{pars_c5}).\footnote{The value of
  norm$_{bb}$ also coincides within 90\% confidence interval if one propagates
  the uncertainty of blank-sky background normalization.}  Also shown in
Table~\ref{pars_c5} are the results of \citet{Takahashi:04}, who
analyzed diffuse emission in the central $6'$ of M31.\footnote{The
appreciable difference
  between our errors and those of \citet{Takahashi:04} is due to the fact that
  we did not fix the metal abundances equal to each other. This was essential for our purposes, because of the
  clear presence of the ``line forest'' at  energies below 2.0 $\keV$ (see
  Sec.~\ref{restrictions} and Fig.~\ref{s_r513}).} %
Below we discuss separately the fitting of ESAS and SBS spectra.

\subsection{ESAS spectra}%
\label{sec:esas}

We build $0.5-10.0\;\kev$ MOS spectra of \texttt{circle5} and
\texttt{ring5-13} regions for 3 observations from
Table~\ref{M31_centre}.\footnote{We exclude the region 1.3--1.8
keV due to the
  presence of two strong unmodelled instrumental lines, see
  Sec.~\ref{analysis_esas}.}  Thus for each spatial region we have 6 spectra
to fit - from observations with MOS1 and MOS2 cameras. We fix the
model parameters to be equal for all six spectra from the same
spatial region (except for normalization of the remaining soft
proton background, as the spectra from different observations are
slightly different).

\begin{table*}
\begin{tabular}{|l|c|c|c|c|c|c|c|c|c|}
  \hline
  Parameters&$kT_{disk},$&norm$_{disk}$&norm$_{bb}$,&$kT_{1}$,&norm$_{1}$,&$kT_{2}$,&norm$_{2}$,&$kT_{3}$,&norm$_{3}$,\\
  &\keV&&10$^{-6}$&\keV&10$^{-3}$&\keV&10$^{-3}$&\keV&10$^{-3}$\\
  \hline
\texttt{circle5},ESAS &$0.722_{-0.103}^{+0.236}$&0.098$_{-0.060}^{+0.098}$&6.71$_{-1.07}^{+1.07}$&0.634$_{-0.059}^{+0.169}$&0.16$_{-0.04}^{+11.55}$&0.396$_{-0.141}^{+0.113}$&0.69$_{-0.31}^{+0.75}$&0.171$_{-0.054}^{+0.031}$&1.08$_{-0.45}^{+63.44}$\\
\hline
\texttt{circle5},SBS&0.549$_{-0.030}^{+0.171}$&0.117$_{-0.047}^{+0.116}$&8.61$_{-0.65}^{+0.92}$&0.640$_{-0.184}^{+0.144}$&0.26$_{-0.26}^{+1.09}$&0.385$_{-0.107}^{+0.069}$&0.60$_{-0.60}^{+0.64}$&0.146$_{-0.122}^{+0.104}$&0.35$_{-0.35}^{+2.65}$\\
\hline
\texttt{ring5-13},ESAS&0.655$_{-0.037}^{+0.192}$&0.249$_{-0.162}^{+0.386}$&43.0$_{-9.3}^{+4.5}$&$0.615_{-0.138}^{+0.121}$&0.53$_{-0.53}^{+0.50}$&0.352$_{-0.118}^{+0.092}$&0.36$_{-0.36}^{+0.63}$&0.102$_{-0.033}^{+0.199}$&10.1$_{-7.7}^{+38.6}$\\
\hline
\texttt{ring5-13},SBS&0.628$_{-0.139}^{+0.229}$&0.126$_{-0.092}^{+0.266}$&25.6$_{-3.2}^{+4.0}$&0.594$_{-0.082}^{+0.160}$&1.25$_{-0.69}^{+34.19}$&0.375$_{-0.087}^{+0.040}$&2.48$_{-1.93}^{+38.4}$&0.155$_{-0.074}^{+0.043}$&10.4$_{-10.4}^{+55.9}$\\
  \hline TOKM,\newline
  EPIC&$0.88^{+0.08}_{-0.07}$&&&$0.61^{+0.03}_{-0.02}$&&
  $0.30^{+0.03}_{-0.02}$&&$0.12^{+0.03}_{-0.02}$&\\
  \hline TOKM,\newline
  ACIS&$0.89^{+0.02}_{-0.01}$&&&$0.60^{+0.03}_{-0.02}$&&
  0.30$^{+0.01}_{-0.01}$&&0.10$^{+0.01}_{-0.01}$&\\
  \hline
\end{tabular}
\caption{Model parameters from regions \texttt{circle5} and
  \texttt{ring5-13}. Also   shown are 90\% confidence
intervals for fitted parameters. Results
  of~\citet{Takahashi:04} ($6'$ circular region in this case) are marked as ``TOKM''.}
\label{pars_c5}
\end{table*}

Since ESAS software subtracts only the instrumental background
component, the remaining cosmic background should be modelled. The
cosmic background component is modelled with the help of
\texttt{Xspec} model \texttt{apec+(apec+pow)*wabs}, according to
the ESAS manual. A cool ($\sim 0.1 \keV$), unabsorbed
\texttt{apec} \citep{Smith:01} component represents the thermal
emission from the Local Hot Bubble. The hot ($\sim 0.25 \keV$),
absorbed \texttt{apec} component represents emission from the
hotter halo and/or intergalactic medium. The last, absorbed
\texttt{pow}erlaw component with powerlaw index $\Gamma = 1.41$
represents the unresolved background from cosmological sources. We
kept its normalization fixed for each region; it corresponds to
$8.88\cdot 10^{-7}$ Xspec units per square arcmin, or to 10.5
photons \keV$^{-1}$ s$^{-1}$ cm$^{-2}$ sr$^{-1}$. The
corresponding hydrogen column density in \texttt{wabs} was left to
vary below its Galactic value $n_{H} = 6.7 \cdot 10^{20} cm^{-2}$
\citep{Morrison:83}. To model the soft proton contamination, we
used \texttt{bknpow/b} model (we fix its break energy at $3.3$
\keV), where index \texttt{/b} means that this component is not
folded through the instrumental effective area (in \texttt{Xspec}
versions 11 and earlier).

The \texttt{diskbb+bbody} (the same as the LMXB model in
\citealt{Takahashi:04}) component describes the point sources, which were not
excluded. We fitted the diffuse M31 component in outer regions with the help
of the sum of three \texttt{vmekal} \citep{Mewe:86,Liedahl:95} models with
fixed temperatures and abundances. The \texttt{wabs} column density was fixed
at its Galactic value.

\subsection{SBS spectra}
\label{xspec-sbs}

We fitted the data from MOS and PN cameras, processed using SBS
method (both separately and combined).  As both cosmic and
instrumental background is subtracted in SBS method, we fitted MOS
and PN spectra on
\texttt{wabs*(diskbb+bbody+vmekal+vmekal+vmekal)} \texttt{Xspec}
model at the energy range 0.6--10.0~keV (0.6--12.0~keV in case of
PN camera).  The reduced $\chi^{2}$ obtained by fitting our
spectra are shown in Table~\ref{chi}; fit parameters are shown in
Table~\ref{pars_c5}.

\begin{figure}
  \includegraphics[width=0.75\columnwidth,angle=270]{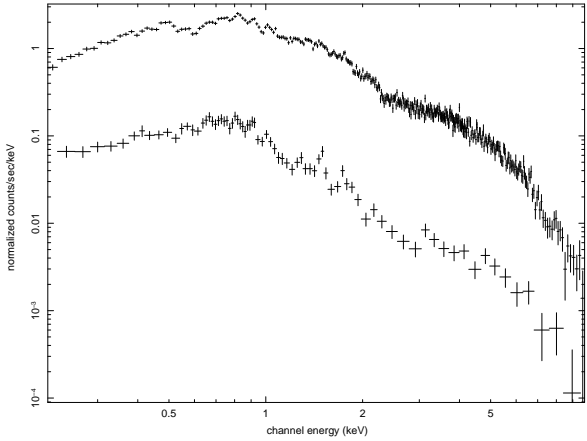}
\caption{\small Folded MOS1 spectra from \texttt{circle5} region,
  ObsID 0112570401, with (top) and without (bottom) point sources.}
\label{c5_s_c5}
\end{figure}

\begin{figure}
\includegraphics[width=0.75\columnwidth,angle=270]{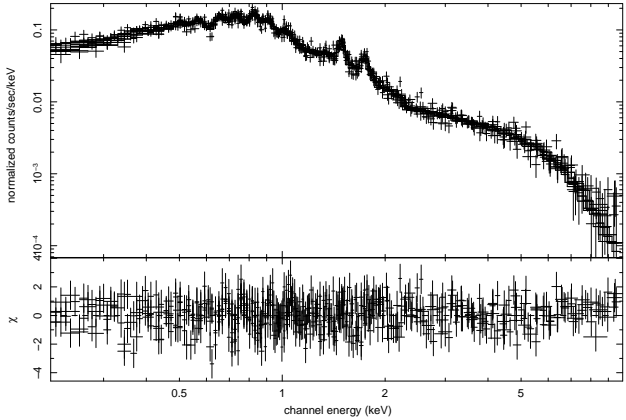}
\caption{\small Folded spectra and best-fit model from
\texttt{circle5} region, with excluded point sources.}
\label{s_c5}
\end{figure}

\begin{table}
{
\begin{tabular}{|l|c|c|}
\hline Region&Reduced $\chi^2$& Number d.o.f.\\
\hline
ESAS, \texttt{circle5}&1.071&399\\
\hline
SBS MOS, \texttt{circle5}&1.102&371\\
\hline
ESAS, \texttt{ring5-13}&1.109&1608\\
\hline
SBS MOS, \texttt{ring5-13}&0.994&1735\\
\hline
SBS PN, \texttt{ring5-13}&1.007&2754\\
\hline
SBS PN-OOT, \texttt{ring5-13}&0.995&2715\\
\hline
SBS MOSPN-OOT, \texttt{ring5-13}&1.009&4082\\
\hline
\end{tabular}
} \caption{Reduced $\chi^{2}$ for our regions.} \label{chi}
\end{table}

\subsection{Producing restrictions on sterile neutrino
  parameters}\label{restrictions}

In this subsection we describe two different techniques of searching for the
narrow (compared to the spectral resolution of \xmm) decay line in the
spectra, processed by ESAS and SBS methods.

\begin{figure*}
  \centering
  \begin{tabular}[c]{cc}
   \includegraphics[width=0.95\columnwidth,angle=0]{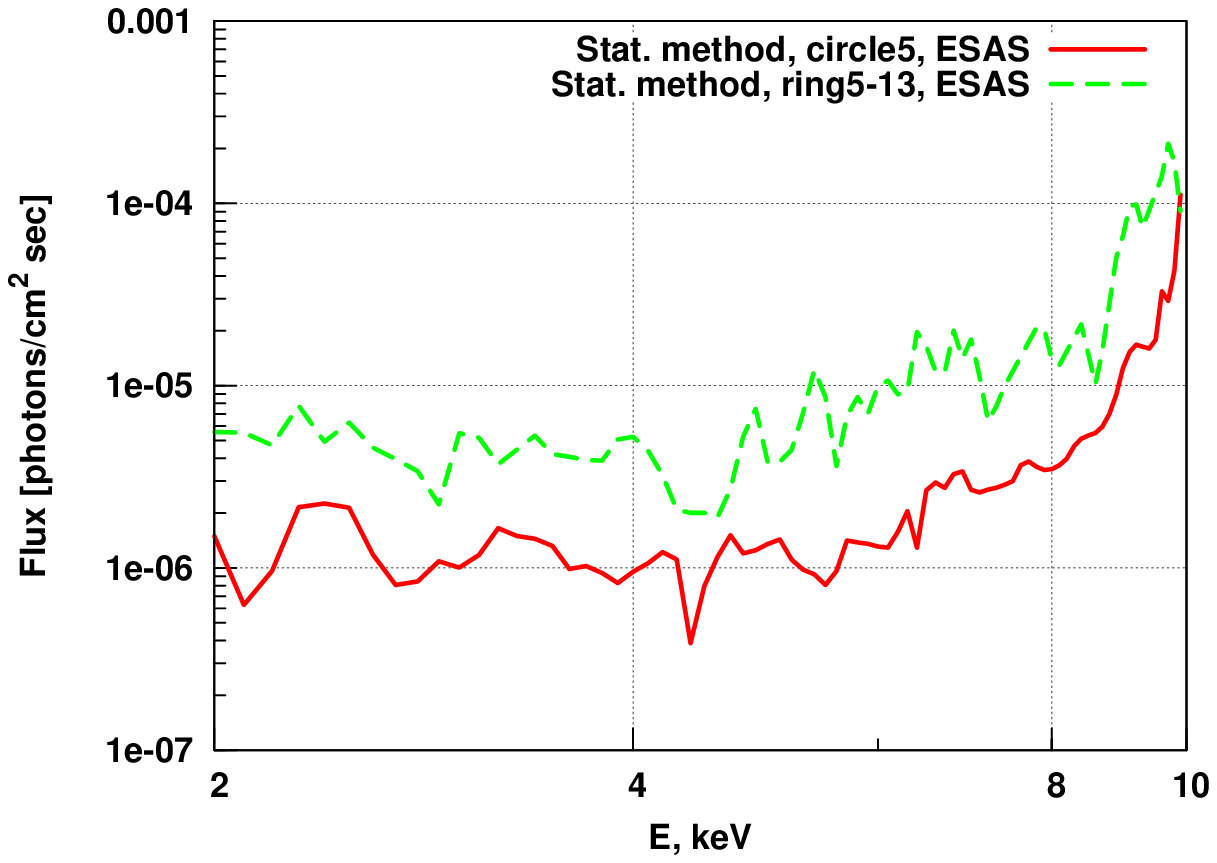}
 &  \includegraphics[width=0.95\columnwidth,angle=0]{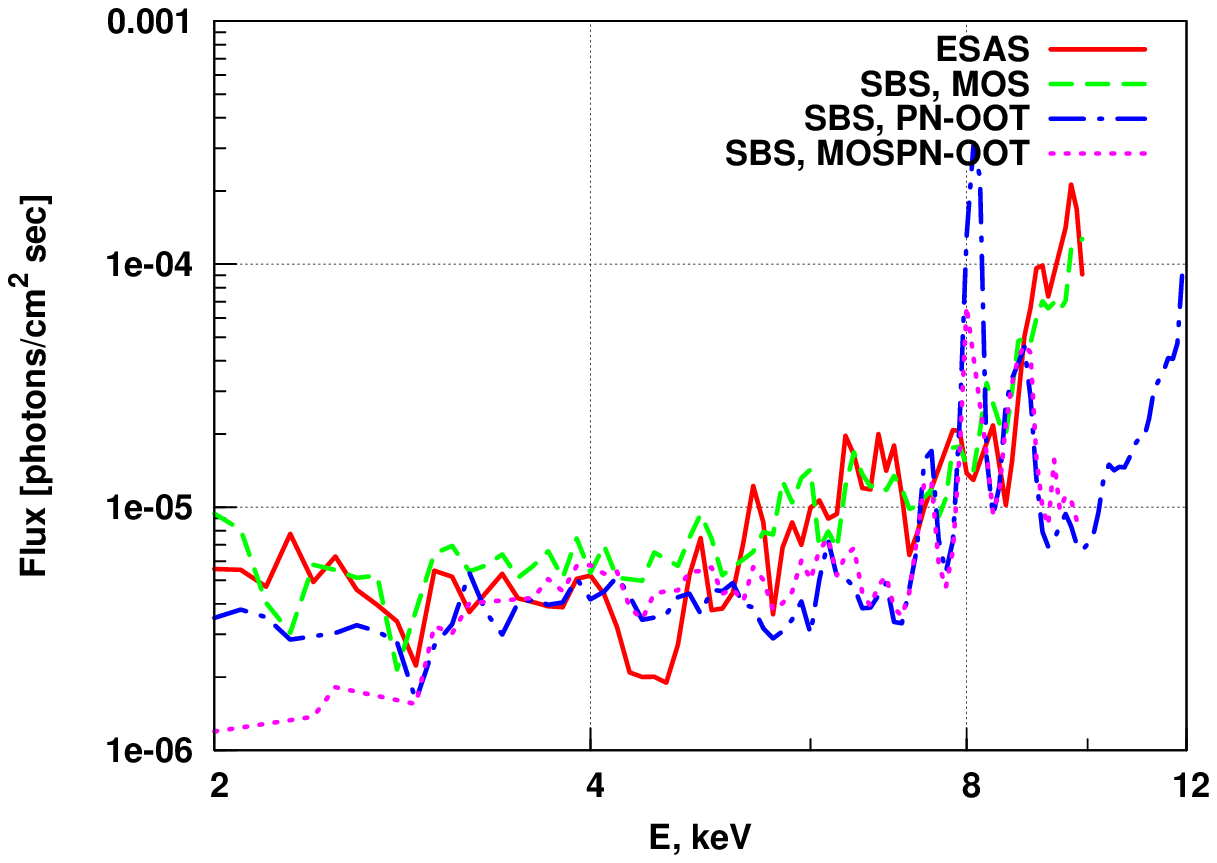}
  \end{tabular}
  \caption{\small $3\sigma$ upper limit on the DM line flux (the region of parameter
    space \textit{above} the curves is excluded). \textit{Left panel:} upper
    limits from the different spatial regions for the spectra, processed by
    ESAS method.  \textit{Right panel:} upper limits for the \texttt{ring5-13}
    region for both ESAS and SBS methods.}
  \label{fig:flux-esas-sbs}
\end{figure*}

As shown on Fig.~\ref{s_r513}, above $2.0\kev$ there are few
emission lines in the model of the spectrum of M31, and continuum
emission dominates. In this case, it is possible to apply the
``statistical'' method, discussed e.g. in \citet{Boyarsky:06c}.
Namely, after fitting the spectra with the selected models
(Secs.~\ref{sec:esas}--\ref{xspec-sbs} above), we add an extra
Gaussian line with the help of \texttt{Xspec} command
\texttt{addcomp}. We then freeze its energy $E_{\gamma}$, leave
the line width $\sigma$ to vary within 0--10 \eV, and repeat the
fit. For each line energy, we re-fit the model and derive an upper
limit on the flux in the Gaussian line, allowing all other model
parameters to vary.  In particular we allow the abundances of
heavy elements, that produce the thermal emission lines to vary.
This produces the most conservative restrictions as the added line
could account for some of the flux from the thermal components.
After that we calculate the $3\sigma$ error with the help of
\texttt{Xspec} command $\texttt{error}\; \langle\texttt{line
  norm}\rangle\; \texttt{9.0}$. To obtain conservative upper limits, we allow
as much freedom as possible for the parameters of the thermal
model. The $3\sigma$ upper limit on the DM line flux is shown in
Fig.~\ref{fig:flux-esas-sbs}. These flux restrictions can be
turned into constraints on parameters of the sterile neutrino
($m_s$ and $\sin^2(2\theta)$), using Eq.~(\ref{flux}) and the
value of the $M_\dm^{fov}$ from the Table~\ref{DM_res2} for the
model~\textbf{M31B}.

Below 2.0 \keV, there are a lot of strong emission lines, which
dominate over the continuum, creating a ``line forest''. As the
intrinsic widths of these lines are much more narrow than the
spectral resolution of EPIC cameras of \xmm, and the abundances of
various elements are known with large uncertainties, it is very
hard to reliably distinguish these emission lines from a possible
DM decay line. Therefore, to produce robust constraints, we apply
the ``full flux'' method below $2\kev$.  In this method, we equate
the DM line flux to the full flux plus 3 flux uncertainties over
the energy interval $\Delta E$ equal to the spectral resolution of
the instrument.\footnote{\label{fn:1}To find the proper value of
$\Delta E$, we
  fold thin Gaussian line with appropriate RMF, and then evaluate FWHM of
  obtained broadened line. The FWHM $\Delta E$, calculated in such a manner,
  slowly increases with line energy and changes from $0.18$~keV to $0.21$~keV
  in the $0.5-2.0$~keV energy region.}

We also produce \textit{model-dependent} ``statistical''
constraints below 2.0 keV. To reduce model uncertainty, we fix
most metal abundances at their values known from optical
observations of M31
\citep{Jacoby:99,Jacoby:86,Dennefeld:81,Blair:82}. The confidence
ranges of these abundances are shown in Table \ref{abundances}.

\begin{table*}
{
\begin{tabular}{|l|c|c|c|c|c|c|c|}
\hline &He&C&N&O&Ne&S&Ar\\
\hline
\citet{Jacoby:99}&1.3$^{+0.3}_{-0.3}$&1.0$^{+0.7}_{-0.4}$&1.1$^{+1.0}_{-0.6}$&0.3$^{+0.2}_{-0.1}$&0.3$^{+0.2}_{-0.1}$&1.5$^{+1.2}_{-0.7}$&0.3$^{+0.2}_{-0.1}$\\
\hline
\citet{Jacoby:86}&1.3$^{+0.4}_{-0.3}$&-&0.5$^{+0.3}_{-0.2}$&0.4$^{+0.1}_{-0.1}$&0.5$^{+0.2}_{-0.2}$&-&-\\
\hline
\citet{Dennefeld:81}&-&0.2&1.0$^{+0.2}_{-0.2}$&0.3$^{+0.1}_{-0.1}$&-&0.8$^{+0.5}_{-0.5}$&-\\
 \hline
\citet{Blair:82}, SNRs&1.6$^{+0.3}_{-0.3}$&-&0.6$^{+0.3}_{-0.3}$&0.4$^{+0.1}_{-0.1}$&0.9$^{+0.1}_{-0.1}$&0.4$^{+0.1}_{-0.1}$&-\\
 \hline
\citet{Blair:82}, HII regions&-&-&0.4$^{+0.3}_{-0.3}$&0.9$^{+0.5}_{-0.5}$&-&0.8$^{+0.5}_{-0.5}$&-\\
 \hline
Our allowed range&1.0..1.9&0.2..1.7&0.1..2.1&0.2..1.4&0.2..1.0&0.3..2.7&0.2..0.5\\
\hline
\end{tabular}
} \caption{Abundances from optical observations (in solar units).
{Our allowed range of abundances, used for construction the
model-dependent restriction (see Sec.~\ref{restrictions}), is also
shown.} } \label{abundances}
\end{table*}

\begin{figure}
  \includegraphics[width=0.95\columnwidth,angle=0]{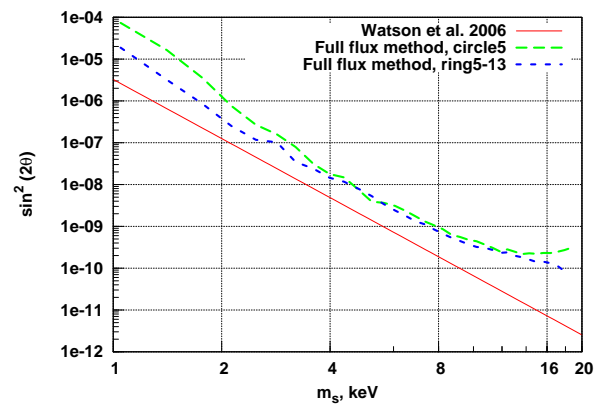}
\caption{\small Our limits on
$\bigl(m_{s},\sin^{2}(2\theta)\bigr)$
  parameters, obtained by using the full flux method from different spatial
  regions of M31 (a region of parameter space \textit{above} a curve is
  excluded). The restriction from W06 is shown for
  comparison.}
\label{rest_ff}
\end{figure}

To compare our results with previous work on M31 \citep[hereafter
W06]{Watson:06} we performed full flux analysis in the whole
region of energies of the MOS camera of \xmm. The results are
shown in Fig.~\ref{rest_ff}. One can see that our full flux
results from \texttt{circle5} region are somewhat weaker that the
corresponding results of W06 (by a factor 2--3 in the region $m_s
\sim 4\kev$; more than an order of magnitude at $m_s \lesssim
2\kev$ and $m_s \gtrsim 12 \kev$).  There are several reasons for
this. As discussed in Sec.~\ref{sec:DM_calc} we use an $\sim 8$
times lower estimate for the DM mass within the FoV, because we
use the more recent and more conservative DM profile
of~\cite{Widrow:05} and compute the amount of DM by explicit
integration over the FoV with removed point sources. At the same
time, comparing our diffuse spectrum
(Figs.~\ref{c5_s_c5}-\ref{s_c5}) with Fig.~1 in~W06, we see that
the intensity of our diffuse spectrum is $\sim 2-3$ times lower
(due to the $\sim 4$ times larger number of point sources
removed). Therefore, one would expect a factor 2--3 difference
between our results (as indeed is seen at $m_s \sim 4\kev$).

An additional discrepancy at low energies is due to the different
choice of the energy bin intervals. In W06 the energy bin interval
was chosen according to the empirical formula $\Delta E =
E_{\gamma}/30 = m_{s}/60$, while we have determined it using the
\xmm\ response matrices (as described in footnote~\ref{fn:1}
above).  The difference is most prominent at low energies: e.g. at
$E\sim 1\kev$ we obtain $\Delta E \approx 0.2\keV$, which is $\sim
6$ times bigger than the value, used by W06.  Therefore, at small
energies we would expect constraints about an order of magnitude
lower than those of~W06, as Fig.~\ref{rest_ff} indeed
demonstrates.

The other important effect, seen in Fig.~\ref{rest_ff}, is the
high-energy behaviour. Our restrictions remain nearly constant for
$m_{s} \gtrsim 12 \keV$ ($E_\gamma \gtrsim 6\kev$), in contrast to
the steeply decreasing results of W06.  This is due to the fact
that W06 used an energy-averaged count-rate-to-flux conversion
factor (i.e., the telescope effective area): see Sec.~IV of~W06.
However, the effective area of the \xmm\ {MOS cameras } declines
sharply with energy, essentially going to zero at
9--10~keV.\footnote{For PN camera
  this happens at $\sim 12$~keV (c.f.  Fig.~\ref{fig:flux-esas-sbs}).}
Therefore, after a proper conversion, a constant count rate at
high energies, assumed by W06 would correspond to a sharply rising
physical flux in $\mathrm{photons/(s\cdot cm^2)}$, which is of
course incorrect. We performed a full data analysis, taking into
account the dependence of the effective area on the energy and our
constraints weaken sharply at high energies. This effect is
well-known and is present in many papers that perform spectral
analysis of \xmm or \emph{Chandra} data.

\begin{figure}
  \includegraphics[width=0.95\columnwidth,angle=0]{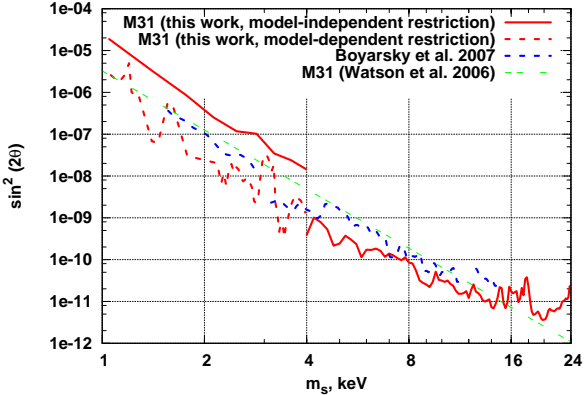}
\caption{\small Restrictions on $(m_{s},\sin^{2}(2\theta))$
  plane. The strongest previous limits of~\citet{Boyarsky:06d} as well as
  results of W06 are shown for comparison. The region \emph{above} the curve is
  excluded. }
\label{final}
\end{figure}

Our final constraints are shown in Fig.~\ref{final}. At masses
$\mdm \ge 4\kev$ (energies $E_\gamma \ge 2\kev$) we use the
results of statistical constraints from the \texttt{ring5-13}
region. To produce the final restriction, we choose, for each
value of $m_{s}$, the \emph{minimal} value of $\sin^{2}(2\theta)$.
For $\mdm < 4\kev$ ($E_\gamma < 2\kev$) we plot both the
model-independent (full flux) and the model-dependent constraints.
The restrictions of~\citet{Boyarsky:06d} and~\citet{Watson:06} are
shown for comparison.

The high-energy behaviour of our final statistical constraints
differs from that of in Fig.~\ref{rest_ff}. There are several
reasons for this. Firstly, in Fig.~\ref{rest_ff} we showed the
full flux restrictions from the MOS camera (to compare our results
with those of W06), while in Fig.~\ref{final} we used the combined
constraints from both MOS and PN cameras. The PN camera has a
wider energy range: its effective area decreases only above
$E\approx 10$~\keV\footnote{\xmm Users Handbook, Sec. 3.2.2.1,
\url{http://xmm.esac.esa.int/external/xmm_user_support/documentation/uhb_2.5}},
which explains the weakening of constraints on Fig.\ref{final} for
$m_s \gtrsim 20\kev$. The ``peak'' at $m_{s} \approx 16-18$~\keV,
is due to the presence of strong Cu instrumental lines in the PN
background spectrum (\citealt{Strueder:01}, see also
Fig.~\ref{fig:flux-esas-sbs}). This region has, thus, higher
errors, which weaken the constraints. Finally, we used several
jointly fitted spectra (up to 9 in MOSPN-OOT dataset) in our
``statistical'' method, as opposed to the restrictions in
Fig.\ref{rest_ff} where we used only one spectrum. The combination
of several spectra improves the bounds, as statistical errors
decrease.

\section{Results and conclusions}%
\label{results}

Using available \xmm data on the central region of the Andromeda
galaxy (M31), we obtained new restrictions on sterile neutrino
Dark Matter parameters.  We analyzed various DM distributions for
the central part of M31, and obtained a conservative estimate of
the DM mass inside the central $13'$, using the model M31B
of~\citet{Widrow:05}. This DM distribution turned out to be the
most conservative among those which studied the DM distribution in
the inner part of M31.\footnote{We would like to notice, however,
that in the work~\cite{Kerins:04}, a number of ``extreme'' (i.e.
maximizing contributions of disk, spheroid or halo) models are
considered. Some of these models would reduce an estimated DM
signal from the inner $13'$ (and correspondingly our limits) by a
factor $\sim 2$. We chose to use the family of models, shown on
  Fig.~\ref{DM_signal}, as they qualitatively agree with each other and do not
  contain any ``extreme'' assumptions. However, below, in deriving a
  model-dependent upper limit of the mass of the DM particle, we will
  introduce an additional penalty factor, to account for this and other
  possible systematic uncertainties.}

We found that exclusion of numerous point
sources from the central part significantly improves our limits, therefore we
have also calculated the DM mass in such ``cheesed'' regions with the help of
Monte Carlo integration.

As the surface brightness is low in the selected regions, the
choice of the background subtraction method is important. We
processed \xmm data from these regions with the help of two
different background subtraction techniques -- the Extended
Sources Analysis Software (ESAS), and the blank-sky background
subtraction (SBS), using the blank-sky background dataset of
\citet{Read:03}. We have shown that these totally different
background subtraction methods give similar results.

To compare our results with the previous work on M31
\citep[\textbf{W06}]{Watson:06}, we obtained the full flux
restriction from the central $5'$ of M31. Our full flux results
(shown in Fig.~\ref{rest_ff}) mostly reproduce the results of~W06,
up to differences arising from our more conservative estimate of
expected DM signal and proper data analysis (see
Sec.~\ref{restrictions} for detailed discussion).

Our final upper limits (both model-dependent and
model-independent) are shown in Fig.~\ref{final}. We improved the
previous bounds of W06 on $\sin^2(2\theta)$ by as much as an order
of magnitude for masses $4\kev\lesssim m_s \lesssim 8\kev$. Due to
the significant low-energy thermal component in M31 diffuse
emission, to produce the model-independent constraints, we have
used the ``full flux'' method for $m_s < 4.0 \kev$ (i.e. $E_\gamma
< 2.0\kev$). In this region, the strongest constraints remain
those of~\cite{Boyarsky:06d}. We have also produced
model-dependent constraints for $E_\gamma < 2.0 \keV$, using the
``statistical'' method; in this case we found the best-fit model
by fixing the metallic abundances at the level of optical
observations.

The comparison of our upper limit with the lower bound on sterile neutrino
pulsar kick mechanism \citep{Fuller:03} improves the previous bounds and can
exclude part of the parameter region (for 4 \keV $< m_{s} <$ 20 \keV).

\begin{figure}
  \centering \includegraphics[width=\linewidth]{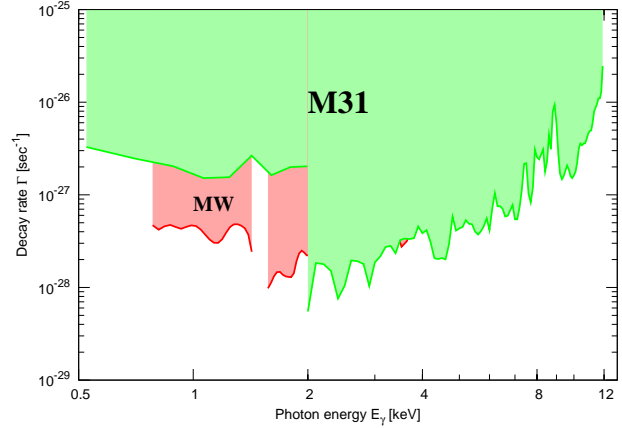}
  \caption{Constraints on the decay width $\Gamma$ of any radiatively decaying
    DM from this work (marked ``M31'') and \protect\citet{Boyarsky:06d}
    (marked ``MW''). The shaded region of parameters is excluded.}
  \label{fig:gamma}
\end{figure}

Finally, it should be noticed that although throughout this paper
we were writing about the sterile neutrino DM, the results of this
work are equally applicable to \emph{any} decaying DM candidate
(e.g. gravitino), emitting photon of energy $E_\gamma$ and having
decay width $\Gamma$. Our final results in this case are presented
in Fig.~\ref{fig:gamma}.  For other works discussing cosmological
and astrophysical effects of decaying DM
see~\cite{DeRujula:80,Berezhiani:87,Doroshkevich:89,Berezhiani:90a,Berezhiani:90b}.
An extensive review of the results can also be found in the book
by~\cite{Khlopov:97}.

\subsection{Sterile neutrino in Dodelson-Widrow model}
\label{sec:ster-neutr-dodels}

\begin{figure}
  \centering
  \includegraphics[width=\linewidth]{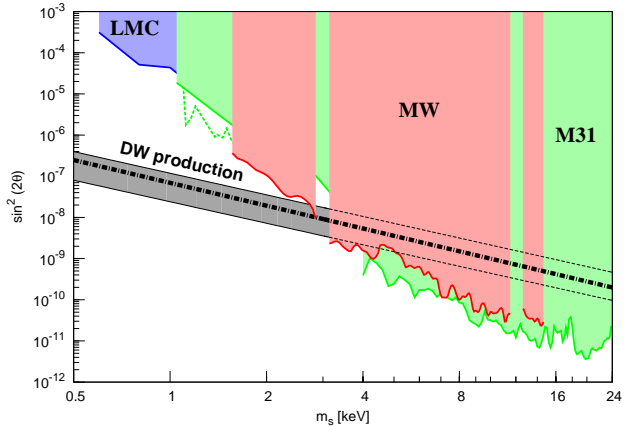} %
  \caption{Current X-ray constraints, combined with the DW production
    model. Colored regions are excluded. The grey region shows the range of
    parameters which give correct abundance in the DW model~\citep{Asaka:06c}.
    The color shaded regions mark the restrictions from
    ``LMC''~\citep{Boyarsky:06c}, ``MW''~\citep{Boyarsky:06d} and ``M31''
    (this work). Model-dependent restrictions from M31 for $m_s < 2\kev$ are
    shown in (green) dashed line.}
  \label{fig:dw}
\end{figure}

The results of this work have important consequences to one of the
production models for the sterile neutrino, the so-called
``Dodelson-Widrow'' (DW) scenario -- production through
(non-resonant) oscillations with an active
neutrino~\citep{Dodelson:93}. The computation of the abundance is
complicated in this case by the fact that the production mainly
happens around the QCD transition and therefore QCD contributions
are hard to compute~\citep[see][and refs.  therein]{Asaka:06b}. A
first-principles computation, taking into account all QCD
contributions in a proper way, was performed in~\citet{Asaka:06c}.

We compare the results of this computation with X-ray bounds
obtained in this work and previous works in Fig.~\ref{fig:dw}. The
upper and lower dashed lines, bounding the grey area, correspond
to the DW production scenario when all hadronic uncertainties are
pushed in one or another direction; the thick central line
corresponds to the most probable relation between $m_s$ and
$\sin^2(2\theta)$. Upon comparison with X-ray bounds, we find that
the upper bound on the DM mass in the DW scenario is reliably
below $m_s < 4 \kev$ (even if we push our X-ray bounds up by a
factor of 2, to account for some yet unknown systematics and push
all the uncertainties in hadronic contributions to the DW
production in one direction).

This improves by 50\% the previous bound $m_s < 6\kev$ of
\citet{Asaka:06c}. Notice that other bounds on $m_s$, that
appeared in the literature (e.g. $m_s < 3.5\kev$ of
\citet{Watson:06} and $m_s < 3\kev$ of~\citet{Boyarsky:06c}) were
based on the computations of~\citet{Abazajian:05a}, which did not
take into account all QCD contributions.

Our present results may be combined with the Lyman-$\alpha$
analysis of~\cite{Seljak:06,Viel:06,Viel:07}. As follows from the
most recent analysis of \cite{Viel:07}, if one uses only the
high-resolution high-redshift Lyman-$\alpha$ spectra
of~\citet{Becker:06} then one finds the lower bound on the sterile
neutrino DM mass in the DW scenario to be $m_s > 5.6\kev$, which
is in contradiction with our current upper bound $m_s < 4 \kev$
(but would have left a narrow allowed window for $m_s$ if one had
used the previous bound $m_s < 6\kev$ of \citealt{Asaka:06c}). If
one takes into account the low-resolution SDSS Lyman-$\alpha$
dataset~\citep{McDonald:05}, used in~\citet{Seljak:06,Viel:06},
this contradiction becomes much stronger. Although the
Lyman-$\alpha$ method relies on a very complicated analysis with
(???) some unknown systematic uncertainties, it seems that the
model in which all of the DM is produces through the DW scenario
is ruled out.

However, there is another way to produce the sterile neutrino
through oscillations with active neutrinos (resonant production in
the presence of lepton asymmetries,~\citealt{Shi:98} (SF)). In
this case, one qualitatively expects that the results of the
Lyman-$\alpha$ analysis can be lowered by a significant amount, as
for the same mass, the mean velocity (free-streaming length) in
the SF model can be much lower than in the DW model.  However, as
sterile neutrinos are produced in the non-equilibrium way and
their spectrum differs significantly from the thermal one, the
actual Lyman-$\alpha$ bounds may depend not only on the
free-streaming but also on the detailed shape of the spectrum. The
detailed analysis of the SF production and corresponding
re-analysis of the Lyman-$\alpha$ data is needed.  Currently, the
SF mechanism is not ruled out.

Finally, there is also the possibility of production of the
sterile neutrino DM through the decay of the light
inflaton~\citep{Shaposhnikov:06}, which cannot be ruled out by
X-ray observations.

Therefore, the sterile neutrino remains a viable and interesting
DM candidate, which can be either warm or cold.  One of the most
interesting ranges of parameters is that of low masses, which is
also in the potential reach of laboratory
experiments~\citep{Bezrukov:06} and will be probed with future
X-ray spectrometers~\citep{Boyarsky:06f,EDGE:07}.\footnote{See
also EDGE
  Project:
  \url{http://projects.iasf-roma.inaf.it/edge}} %
However, the search for the sterile neutrino DM signal in all energy ranges
above Tremaine-Gunn limit should also be conducted.

\section*{Acknowledgements}

We would like to thank B.~Gripaios, A.~Neronov, J.~Nevalainen,
M.~Markevich, M.~Shaposhnikov, C.~Watson for useful comments.
D.I. is grateful to ESAC team and especially to M.~Kirsch, for
granting his stay at ESAC and for useful discussions. D.I.  and
V.S. are also grateful to M.~Ehle, R.~Saxton and S.~Snowden for
useful discussions about ESAS software, to Scientific and
Educational Centre\footnote{\url{http://sec.bitp.kiev.ua}} of the
Bogolyubov Institute for Theoretical Physics in Kiev, Ukraine, and
especially to V.~Shadura, for creating wonderful atmosphere for
young Ukrainian scientists, and to Ukrainian Virtual Roentgen and
Gamma-Ray Observatory
VIRGO.UA\footnote{\url{http://virgo.bitp.kiev.ua}} and computing
cluster of Bogolyubov Institute for Theoretical
Physics\footnote{\url{http://grid.bitp.kiev.ua}}, for using their
computing resources. This work was supported by the Swiss National
Science Foundation and the Swiss Agency for Development and
Cooperation in the framework of the programme SCOPES - Scientific
co-operation between Eastern Europe and Switzerland. D.I. also
acknowledges support from the INTAS project No. 05-1000008-7865.
The work of A.B. was (partially) supported by the EU 6th Framework
Marie Curie Research and Training network "UniverseNet" (MRTN-
CT-2006-035863). O.R. would like to acknowledge support of the
Swiss Science Foundation.

\let\jnlstyle=\rm\def\jref#1{{\jnlstyle#1}}\def\aj{\jref{AJ}}
  \def\araa{\jref{ARA\&A}} \def\apj{\jref{ApJ}} \def\apjl{\jref{ApJ}}
  \def\apjs{\jref{ApJS}} \def\ao{\jref{Appl.~Opt.}} \def\apss{\jref{Ap\&SS}}
  \def\aap{\jref{A\&A}} \def\aapr{\jref{A\&A~Rev.}} \def\aaps{\jref{A\&AS}}
  \def\azh{\jref{AZh}} \def\baas{\jref{BAAS}} \def\jrasc{\jref{JRASC}}
  \def\memras{\jref{MmRAS}} \def\mnras{\jref{MNRAS}}
  \def\pra{\jref{Phys.~Rev.~A}} \def\prb{\jref{Phys.~Rev.~B}}
  \def\prc{\jref{Phys.~Rev.~C}} \def\prd{\jref{Phys.~Rev.~D}}
  \def\pre{\jref{Phys.~Rev.~E}} \def\prl{\jref{Phys.~Rev.~Lett.}}
  \def\pasp{\jref{PASP}} \def\pasj{\jref{PASJ}} \def\qjras{\jref{QJRAS}}
  \def\skytel{\jref{S\&T}} \def\solphys{\jref{Sol.~Phys.}}
  \def\sovast{\jref{Soviet~Ast.}} \def\ssr{\jref{Space~Sci.~Rev.}}
  \def\zap{\jref{ZAp}} \def\nat{\jref{Nature}} \def\iaucirc{\jref{IAU~Circ.}}
  \def\aplett{\jref{Astrophys.~Lett.}}
  \def\apspr{\jref{Astrophys.~Space~Phys.~Res.}}
  \def\bain{\jref{Bull.~Astron.~Inst.~Netherlands}}
  \def\fcp{\jref{Fund.~Cosmic~Phys.}} \def\gca{\jref{Geochim.~Cosmochim.~Acta}}
  \def\grl{\jref{Geophys.~Res.~Lett.}} \def\jcp{\jref{J.~Chem.~Phys.}}
  \def\jgr{\jref{J.~Geophys.~Res.}}
  \def\jqsrt{\jref{J.~Quant.~Spec.~Radiat.~Transf.}}
  \def\memsai{\jref{Mem.~Soc.~Astron.~Italiana}}
  \def\nphysa{\jref{Nucl.~Phys.~A}} \def\physrep{\jref{Phys.~Rep.}}
  \def\physscr{\jref{Phys.~Scr}} \def\planss{\jref{Planet.~Space~Sci.}}
  \def\procspie{\jref{Proc.~SPIE}} \let\astap=\aap \let\apjlett=\apjl
  \let\apjsupp=\apjs \let\applopt=\ao

\label{lastpage} %

\begin{thebibliography}{111}
\expandafter\ifx\csname natexlab\endcsname\relax\def\natexlab#1{#1}\fi

\bibitem[{Abazajian(2006)}]{Abazajian:05a}
Abazajian, K. 2006, Phys. Rev., D73, 063506, \eprint{astro-ph/0511630}

\bibitem[{Abazajian {et~al.}(2001)Abazajian, Fuller, \& Patel}]{Abazajian:01a}
Abazajian, K., Fuller, G.~M., \& Patel, M. 2001, \prd, 64, 023501,
  \eprint{astro-ph/0101524}

\bibitem[{{Abazajian} {et~al.}(2007){Abazajian}, {Markevitch}, {Koushiappas},
  \& {Hickox}}]{Abazajian:06b}
{Abazajian}, K.~N., {Markevitch}, M., {Koushiappas}, S.~M., \& {Hickox}, R.~C.
  2007, \prd, 75, 063511,
  \adsurl{http://adsabs.harvard.edu/cgi-bin/nph-bib_query?bibcode=2006astro.ph%
.11144A&db_key=PRE}, \eprint{arXiv:astro-ph/0611144}

\bibitem[{{Alcock} {et~al.}(2000)}]{MACHO:00}
{Alcock}, C., {et~al.} 2000, \apj, 541, 270,
  \adsurl{http://adsabs.harvard.edu/abs/2000ApJ...541..270A}

\bibitem[{Arnaud(1996)}]{Arnaud:96}
Arnaud, K.~A. 1996, in A.S.P. Conference Serie, Vol. 101, Astronomical Data
  Analysis Software and Systems V, ed. G.~H. Jacoby \& J.~Barnes (San
  Francisco, ASP), 17

\bibitem[{Asaka {et~al.}(2005)Asaka, Blanchet, \& Shaposhnikov}]{Asaka:05a}
Asaka, T., Blanchet, S., \& Shaposhnikov, M. 2005, Phys. Lett., B631, 151,
  \eprint{hep-ph/0503065}

\bibitem[{Asaka {et~al.}(2006)Asaka, Laine, \& Shaposhnikov}]{Asaka:06b}
Asaka, T., Laine, M., \& Shaposhnikov, M. 2006, JHEP, 06, 053,
  \eprint{hep-ph/0605209}

\bibitem[{Asaka {et~al.}(2007)Asaka, Laine, \& Shaposhnikov}]{Asaka:06c} ---.
  2007, JHEP, 01, 091, \eprint{hep-ph/0612182}

\bibitem[{Asaka \& Shaposhnikov(2005)}]{Asaka:05b}
Asaka, T., \& Shaposhnikov, M. 2005, Phys. Lett., B620, 17,
  \eprint{hep-ph/0505013}

\bibitem[{{Avila-Reese} {et~al.}(2001){Avila-Reese}, {Col{\'{\i}}n},
  {Valenzuela}, {D'Onghia}, \& {Firmani}}]{Avila-Reese:01}
{Avila-Reese}, V., {Col{\'{\i}}n}, P., {Valenzuela}, O., {D'Onghia}, E., \&
  {Firmani}, C. 2001, \apj, 559, 516,
  \adsurl{http://adsabs.harvard.edu/abs/2001ApJ...559..516A},
  \eprint{arXiv:astro-ph/0010525}

\bibitem[{{Baltz} \& {Murayama}(2003)}]{Baltz:01}
{Baltz}, E.~A., \& {Murayama}, H. 2003, JHEP, 5, 67,
  \adsurl{http://adsabs.harvard.edu/abs/2003JHEP...05..067B},
  \eprint{arXiv:astro-ph/0108172}

\bibitem[{Barger {et~al.}(1995)Barger, Phillips, \& Sarkar}]{Barger:95}
Barger, V.~D., Phillips, R. J.~N., \& Sarkar, S. 1995, Phys. Lett., B352, 365,
  \eprint{hep-ph/9503295}

\bibitem[{{Becker} {et~al.}(2007){Becker}, {Rauch}, \& {Sargent}}]{Becker:06}
{Becker}, G.~D., {Rauch}, M., \& {Sargent}, W.~L.~W. 2007, \apj, 662, 72,
  \adsurl{http://adsabs.harvard.edu/abs/2007ApJ...662...72B},
  \eprint{arXiv:astro-ph/0607633}

\bibitem[{Berezhiani \& Khlopov(1990)}]{Berezhiani:90b}
Berezhiani, Z.~G., \& Khlopov, M.~Y. 1990, Sov. J. Nucl. Phys., 52, 60

\bibitem[{Berezhiani {et~al.}(1987)Berezhiani, Vysotsky, \&
  Khlopov}]{Berezhiani:87}
Berezhiani, Z.~G., Vysotsky, M.~I., \& Khlopov, M.~Y. 1987, Sov. J. Nucl.
  Phys., 45, 1065

\bibitem[{Berezhiani {et~al.}(1990)Berezhiani, Vysotsky, Yurov, Doroshkevich,
  \& Khlopov}]{Berezhiani:90a}
Berezhiani, Z.~G., Vysotsky, M.~I., Yurov, V.~P., Doroshkevich, A.~G., \&
  Khlopov, M.~Y. 1990, Sov. J. Nucl. Phys., 51, 1020

\bibitem[{{Bezrukov} \& {Shaposhnikov}(2007)}]{Bezrukov:06}
{Bezrukov}, F., \& {Shaposhnikov}, M. 2007, \prd, 75, 053005,
  \adsurl{http://adsabs.harvard.edu/abs/2007PhRvD..75e3005B},
  \eprint{arXiv:hep-ph/0611352}

\bibitem[{Biermann \& Kusenko(2006)}]{Biermann:06}
Biermann, P.~L., \& Kusenko, A. 2006, Phys. Rev. Lett., 96, 091301,
  \eprint{astro-ph/0601004}

\bibitem[{{Bisnovatyi-Kogan}(1980)}]{Bisnovatyi:80}
{Bisnovatyi-Kogan}, G.~S. 1980, \azh, 57, 899,
  \adsurl{http://adsabs.harvard.edu/abs/1980AZh....57..899B}

\bibitem[{{Blair} {et~al.}(1982){Blair}, {Kirshner}, \& {Chevalier}}]{Blair:82}
{Blair}, W.~P., {Kirshner}, R.~P., \& {Chevalier}, R.~A. 1982, \apj, 254, 50,
  \adsurl{http://adsabs.harvard.edu/abs/1982ApJ...254...50B}

\bibitem[{Bode {et~al.}(2001)Bode, Ostriker, \& Turok}]{Bode:00}
Bode, P., Ostriker, J.~P., \& Turok, N. 2001, \apj, 556, 93,
  \eprint{astro-ph/0010389}

\bibitem[{{Bond} {et~al.}(1980){Bond}, {Efstathiou}, \& {Silk}}]{Bond:80}
{Bond}, J.~R., {Efstathiou}, G., \& {Silk}, J. 1980, Phys. Rev. Lett., 45,
  1980, \adsurl{http://adsabs.harvard.edu/abs/1980PhRvL..45.1980B}

\bibitem[{{Bond} \& {Szalay}(1983)}]{Bond:83}
{Bond}, J.~R., \& {Szalay}, A.~S. 1983, \apj, 274, 443,
  \adsurl{http://adsabs.harvard.edu/abs/1983ApJ...274..443B}

\bibitem[{Boyarsky {et~al.}(2006{\natexlab{a}})Boyarsky, den Herder, Neronov,
  \& Ruchayskiy}]{Boyarsky:06f}
Boyarsky, A., den Herder, J.~W., Neronov, A., \& Ruchayskiy, O.
  2006{\natexlab{a}}, To appear in Astropart. Phys.,
  \adsurl{http://adsabs.harvard.edu/cgi-bin/nph-bib_query?bibcode=2006astro.ph%
.12219B&db_key=PRE}, \eprint{astro-ph/0612219}

\bibitem[{Boyarsky {et~al.}(2006{\natexlab{b}})Boyarsky, Neronov, Ruchayskiy,
  \& Shaposhnikov}]{Boyarsky:05}
Boyarsky, A., Neronov, A., Ruchayskiy, O., \& Shaposhnikov, M.
  2006{\natexlab{b}}, \mnras, 370, 213,
  \adsurl{http://adsabs.harvard.edu/cgi-bin/nph-bib_query?bibcode=2006MNRAS.37%
0..213B&db_key=AST}, \eprint{astro-ph/0512509}

\bibitem[{Boyarsky {et~al.}(2006{\natexlab{c}})Boyarsky, Neronov, Ruchayskiy,
  \& Shaposhnikov}]{Boyarsky:06b}
---. 2006{\natexlab{c}}, \prd, 74, 103506, \eprint{astro-ph/0603368}

\bibitem[{Boyarsky {et~al.}(2006{\natexlab{d}})Boyarsky, Neronov, Ruchayskiy,
  Shaposhnikov, \& Tkachev}]{Boyarsky:06c}
Boyarsky, A., Neronov, A., Ruchayskiy, O., Shaposhnikov, M., \& Tkachev, I.
  2006{\natexlab{d}}, \prl, 97, 261302, \eprint{astro-ph/0603660}

\bibitem[{Boyarsky {et~al.}(2007)Boyarsky, Nevalainen, \&
  Ruchayskiy}]{Boyarsky:06d}
Boyarsky, A., Nevalainen, J., \& Ruchayskiy, O. 2007, \aap, 471, 51,
  \adsurl{http://adsabs.harvard.edu/abs/2007A\%26A...471...51B},
  \eprint{astro-ph/0610961}

\bibitem[{Boyarsky {et~al.}(2006{\natexlab{e}})Boyarsky, Ruchayskiy, \&
  Markevitch}]{Boyarsky:06e}
Boyarsky, A., Ruchayskiy, O., \& Markevitch, M. 2008, \apj, 673,
752,
  \adsurl{http://adsabs.harvard.edu/cgi-bin/nph-bib_query?bibcode=2006astro.ph%
.11168B&db_key=PRE}, \eprint{astro-ph/0611168}

\bibitem[{{Burkert}(1995)}]{Burkert:95}
{Burkert}, A. 1995, \apjl, 447, L25+,
  \adsurl{http://adsabs.harvard.edu/abs/1995ApJ...447L..25B},
  \eprint{arXiv:astro-ph/9504041}

\bibitem[{Carignan {et~al.}(2006)Carignan, Chemin, Huchtmeier, \&
  Lockman}]{Carignan:06}
Carignan, C., Chemin, L., Huchtmeier, W.~K., \& Lockman, F.~J. 2006, \apj, 641,
  L109, \eprint{astro-ph/0603143}

\bibitem[{{Cembranos} {et~al.}(2006){Cembranos}, {Feng}, {Rajaraman}, {Smith},
  \& {Takayama}}]{Cembranos:06}
{Cembranos}, J.~A.~R., {Feng}, J.~L., {Rajaraman}, A., {Smith}, B.~T., \&
  {Takayama}, F. 2006,
  \adsurl{http://adsabs.harvard.edu/abs/2006hep.ph....3067C},
  \eprint{hep-ph/0603067}

\bibitem[{Dalcanton \& Hogan(2001)}]{Dalcanton:00}
Dalcanton, J.~J., \& Hogan, C.~J. 2001, \apj, 561, 35,
  \eprint{astro-ph/0004381}

\bibitem[{{Dar}(1995)}]{Dar:95}
{Dar}, A. 1995, \apj, 449, 550,
  \adsurl{http://adsabs.harvard.edu/abs/1995ApJ...449..550D},
  \eprint{arXiv:astro-ph/9504082}

\bibitem[{{de Rujula} \& {Glashow}(1980)}]{DeRujula:80}
{de Rujula}, A., \& {Glashow}, S.~L. 1980, \prl, 45, 942,
  \adsurl{http://adsabs.harvard.edu/cgi-bin/nph-bib_query?bibcode=1980PhRvL..4%
5..942D&db_key=AST}

\bibitem[{den Herder {et~al.}(2007)}]{EDGE:07}
den Herder, J.~W., {et~al.} 2007, in \procspie, Vol. 6688, 4

\bibitem[{{Dennefeld} \& {Kunth}(1981)}]{Dennefeld:81}
{Dennefeld}, M., \& {Kunth}, D. 1981, \aj, 86, 989,
  \adsurl{http://adsabs.harvard.edu/abs/1981AJ.....86..989D}

\bibitem[{Diemand {et~al.}(2007)Diemand, Kuhlen, \& Madau}]{Diemand:06}
Diemand, J., Kuhlen, M., \& Madau, P. 2007, \apj, 657, 262,
  \adsurl{http://adsabs.harvard.edu/abs/2007ApJ...657..262D},
  \eprint{astro-ph/0611370}

\bibitem[{Dodelson \& Widrow(1994)}]{Dodelson:93}
Dodelson, S., \& Widrow, L.~M. 1994, Phys. Rev. Lett., 72, 17,
  \eprint{hep-ph/9303287}

\bibitem[{Dolgov \& Hansen(2002)}]{Dolgov:00}
Dolgov, A.~D., \& Hansen, S.~H. 2002, Astropart. Phys., 16, 339,
  \eprint{hep-ph/0009083}

\bibitem[{{Doroshkevich} {et~al.}(1989){Doroshkevich}, {Khlopov}, \&
  {Klypin}}]{Doroshkevich:89}
{Doroshkevich}, A.~G., {Khlopov}, M.~I., \& {Klypin}, A.~A. 1989, \mnras, 239,
  923,
  \adsurl{http://adsabs.harvard.edu/cgi-bin/nph-bib_query?bibcode=1989MNRAS.23%
9..923D&db_key=AST}

\bibitem[{{Doroshkevich} {et~al.}(1981){Doroshkevich}, {Khlopov}, {Sunyaev},
  {Szalay}, \& {Zeldovich}}]{Doroshkevich:81}
{Doroshkevich}, A.~G., {Khlopov}, M.~I., {Sunyaev}, R.~A., {Szalay}, A.~S., \&
  {Zeldovich}, I.~B. 1981, New York Academy Sciences Annals, 375, 32,
  \adsurl{http://adsabs.harvard.edu/abs/1981NYASA.375...32D}

\bibitem[{{Einasto} {et~al.}(1974){Einasto}, {Jaaniste}, {J{\^o}eveer},
  {Kaasik}, {Kalamees}, {Saar}, {Tago}, {Traat}, {Vennik}, \&
  {Chernin}}]{Einasto:74}
{Einasto}, J. {et~al.} 1974, Tartu Astrofuusika Observatoorium Teated, 48, 3,
  \adsurl{http://adsabs.harvard.edu/abs/1974TarOT..48....3E}


\bibitem[{{Fogli} {et~al.}(2006){Fogli}, {Lisi}, {Marrone}, {Palazzo}, \&
  {Rotunno}}]{Fogli:05}
{Fogli}, G.~L., {Lisi}, E., {Marrone}, A., {Palazzo}, A., \& {Rotunno}, A.~M.
  2006, Prog. Part. Nucl. Phys., 57, 71,
  \adsurl{http://adsabs.harvard.edu/abs/2006PrPNP..57...71F},
  \eprint{arXiv:hep-ph/0506083}

\bibitem[{Fuller {et~al.}(2003)Fuller, Kusenko, Mocioiu, \&
  Pascoli}]{Fuller:03}
Fuller, G.~M., Kusenko, A., Mocioiu, I., \& Pascoli, S. 2003, Phys. Rev., D68,
  103002, \eprint{astro-ph/0307267}

\bibitem[{{Gates} {et~al.}(1995){Gates}, {Gyuk}, \& {Turner}}]{Gates:95}
{Gates}, E.~I., {Gyuk}, G., \& {Turner}, M.~S. 1995, \apjl, 449, L123+,
  \adsurl{http://adsabs.harvard.edu/cgi-bin/nph-bib_query?bibcode=1995ApJ...44%
9L.123G&db_key=AST}, \eprint{astro-ph/9505039}

\bibitem[{{Geehan} {et~al.}(2006){Geehan}, {Fardal}, {Babul}, \&
  {Guhathakurta}}]{Geehan:06}
{Geehan}, J.~J., {Fardal}, M.~A., {Babul}, A., \& {Guhathakurta}, P. 2006,
  \mnras, 366, 996, \adsurl{http://adsabs.harvard.edu/abs/2006MNRAS.366..996G},
  \eprint{arXiv:astro-ph/0501240}

\bibitem[{{Ghigna} {et~al.}(2000){Ghigna}, {Moore}, {Governato}, {Lake},
  {Quinn}, \& {Stadel}}]{Ghigna:99}
{Ghigna}, S., {Moore}, B., {Governato}, F., {Lake}, G., {Quinn}, T., \&
  {Stadel}, J. 2000, \apj, 544, 616,
  \adsurl{http://adsabs.harvard.edu/cgi-bin/nph-bib_query?bibcode=2000ApJ...54%
4..616G&db_key=AST}, \eprint{astro-ph/9910166}

\bibitem[{{Gilmore}(2007)}]{Gilmore:07b}
{Gilmore}, G. 2007, \adsurl{http://adsabs.harvard.edu/abs/2007astro.ph..3370G},
  \eprint{astro-ph/0703370}

\bibitem[{{Gilmore} {et~al.}(2006){Gilmore}, {Wilkinson}, {Kleyna}, {Koch},
  {Wyn Evans}, {Wyse}, \& {Grebel}}]{Gilmore:06}
{Gilmore}, G., {Wilkinson}, M., {Kleyna}, J., {Koch}, A., {Wyn Evans}, N.,
  {Wyse}, R.~F.~G., \& {Grebel}, E.~K. 2006,
  \adsurl{http://adsabs.harvard.edu/cgi-bin/nph-bib_query?bibcode=2006astro.ph%
..8528G&db_key=PRE}, \eprint{astro-ph/0608528}

\bibitem[{{Gilmore} {et~al.}(2007){Gilmore}, {Wilkinson}, {Wyse}, {Kleyna},
  {Koch}, {Evans}, \& {Grebel}}]{Gilmore:07a}
{Gilmore}, G., {Wilkinson}, M.~I., {Wyse}, R.~F.~G., {Kleyna}, J.~T., {Koch},
  A., {Evans}, N.~W., \& {Grebel}, E.~K. 2007, \apj, 663, 948,
  \adsurl{http://adsabs.harvard.edu/abs/2007ApJ...663..948G},
  \eprint{arXiv:astro-ph/0703308}

\bibitem[{Giunti(2007)}]{Giunti:06}
Giunti, C. 2007, Nucl. Phys. Proc. Suppl., 169, 309, \eprint{hep-ph/0611125}

\bibitem[{Goerdt {et~al.}(2006)Goerdt, Moore, Read, Stadel, \&
  Zemp}]{Goerdt:06}
Goerdt, T., Moore, B., Read, J.~I., Stadel, J., \& Zemp, M. 2006, \mnras, 368,
  1073,
  \adsurl{http://adsabs.harvard.edu/cgi-bin/nph-bib_query?bibcode=2006MNRAS.36%
8.1073G&db_key=AST}, \eprint{astro-ph/0601404}

\bibitem[{Hidaka \& Fuller(2006)}]{Hidaka:06}
Hidaka, J., \& Fuller, G.~M. 2006, \prd, 74, 125015,
  \adsurl{http://adsabs.harvard.edu/abs/2006PhRvD..74l5015H},
  \eprint{astro-ph/0609425}

\bibitem[{{Hidaka} \& {Fuller}(2007)}]{Hidaka:07}
{Hidaka}, J., \& {Fuller}, G.~M. 2007, 706,
  \adsurl{http://adsabs.harvard.edu/abs/2007arXiv0706.3886H},
  \eprint{0706.3886}

\bibitem[{{Hui} {et~al.}(1997){Hui}, {Gnedin}, \& {Zhang}}]{Hui:97}
{Hui}, L., {Gnedin}, N.~Y., \& {Zhang}, Y. 1997, \apj, 486, 599,
  \adsurl{http://adsabs.harvard.edu/cgi-bin/nph-bib_query?bibcode=1997ApJ...48%
6..599H&db_key=AST}, \eprint{astro-ph/9608157}

\bibitem[{{Jacoby} \& {Ciardullo}(1999)}]{Jacoby:99}
{Jacoby}, G.~H., \& {Ciardullo}, R. 1999, \apj, 515, 169,
  \adsurl{http://adsabs.harvard.edu/abs/1999ApJ...515..169J},
  \eprint{arXiv:astro-ph/9812165}

\bibitem[{{Jacoby} \& {Ford}(1986)}]{Jacoby:86}
{Jacoby}, G.~H., \& {Ford}, H.~C. 1986, \apj, 304, 490,
  \adsurl{http://adsabs.harvard.edu/abs/1986ApJ...304..490J}

\bibitem[{{Jansen} {et~al.}(2001){Jansen}, {Lumb}, {Altieri}, {Clavel}, {Ehle},
  {Erd}, {Gabriel}, {Guainazzi}, {Gondoin}, {Much}, {Munoz}, {Santos},
  {Schartel}, {Texier}, \& {Vacanti}}]{Jansen:01}
{Jansen}, F. {et~al.} 2001, \aap, 365, L1

\bibitem[Kerins et al.(2001)]{Kerins:01} Kerins, E., et al.\
2001, \mnras, 323, 13,
\adsurl{http://adsabs.harvard.edu/abs/2001MNRAS.323...13K},
\eprint{arXiv:astro-ph/0002256}

\bibitem[Kerins(2004)]{Kerins:04} Kerins, E.\ 2004, \mnras, 347,
1033, \adsurl{http://adsabs.harvard.edu/abs/2004MNRAS.347.1033K},
\eprint{arXiv:astro-ph/0310537}

\bibitem[{{Khlopov}(1997)}]{Khlopov:97}
{Khlopov}, M.~Y. 1997, Cosmoparticle Physics (World Scientific Pub Co Inc)

\bibitem[{{King}(1962)}]{King:62}
{King}, I. 1962, \aj, 67, 471,
  \adsurl{http://adsabs.harvard.edu/abs/1962AJ.....67..471K}

\bibitem[{{Kleyna} {et~al.}(2003{\natexlab{a}}){Kleyna}, {Wilkinson},
  {Gilmore}, \& {Evans}}]{Kleyna:03}
{Kleyna}, J.~T., {Wilkinson}, M.~I., {Gilmore}, G., \& {Evans}, N.~W.
  2003{\natexlab{a}}, \apjl, 588, L21,
  \adsurl{http://adsabs.harvard.edu/cgi-bin/nph-bib_query?bibcode=2003ApJ...58%
8L..21K&db_key=AST}, \eprint{astro-ph/0304093}

\bibitem[{{Kleyna} {et~al.}(2003{\natexlab{b}}){Kleyna}, {Wilkinson},
  {Gilmore}, \& {Evans}}]{Kleyna:03b}
---. 2003{\natexlab{b}}, \apjl, 589, L59,
  \adsurl{http://adsabs.harvard.edu/cgi-bin/nph-bib_query?bibcode=2003ApJ...58%
9L..59K&db_key=AST}

\bibitem[{{Klypin} {et~al.}(1999){Klypin}, {Kravtsov}, {Valenzuela}, \&
  {Prada}}]{Klypin:99}
{Klypin}, A., {Kravtsov}, A.~V., {Valenzuela}, O., \& {Prada}, F. 1999, \apj,
  522, 82, \adsurl{http://adsabs.harvard.edu/abs/1999ApJ...522...82K},
  \eprint{arXiv:astro-ph/9901240}

\bibitem[{{Klypin} {et~al.}(2002){Klypin}, {Zhao}, \& {Somerville}}]{Klypin:02}
{Klypin}, A., {Zhao}, H., \& {Somerville}, R.~S. 2002, \apj, 573, 597,
  \adsurl{http://adsabs.harvard.edu/cgi-bin/nph-bib_query?bibcode=2002ApJ...57%
3..597K&db_key=AST}, \eprint{astro-ph/0110390}

\bibitem[{{Koposov} {et~al.}(2007){Koposov}, {Belokurov}, {Evans}, {Hewett},
  {Irwin}, {Gilmore}, {Zucker}, {Rix}, {Fellhauer}, {Bell}, \&
  {Glushkova}}]{Koposov:07}
{Koposov}, S. {et~al.} 2007, \apj, 663, 948,
  \adsurl{http://adsabs.harvard.edu/abs/2007arXiv0706.2687K},
  \eprint{0706.2687}

\bibitem[{Kusenko(2006)}]{Kusenko:06a}
Kusenko, A. 2006, Phys. Rev. Lett., 97, 241301, \eprint{hep-ph/0609081}

\bibitem[{{Lasserre} {et~al.}(2000)}]{EROS:00}
{Lasserre}, T., {et~al.} 2000, \aap, 355, L39,
  \adsurl{http://adsabs.harvard.edu/abs/2000A\%26A...355L..39L},
  \eprint{arXiv:astro-ph/0002253}

\bibitem[{{Liedahl} {et~al.}(1995){Liedahl}, {Osterheld}, \&
  {Goldstein}}]{Liedahl:95}
{Liedahl}, D.~A., {Osterheld}, A.~L., \& {Goldstein}, W.~H. 1995, \apjl, 438,
  L115,
  \adsurl{http://adsabs.harvard.edu/cgi-bin/nph-bib_query?bibcode=1995ApJ...43%
8L.115L&db_key=AST}

\bibitem[{{McDonald} {et~al.}(2006){McDonald}, {Seljak}, {Burles}, {Schlegel},
  {Weinberg}, {Cen}, {Shih}, {Schaye}, {Schneider}, {Bahcall}, {Briggs},
  {Brinkmann}, {Brunner}, {Fukugita}, {Gunn}, {Ivezi{\'c}}, {Kent}, {Lupton},
  \& {Vanden Berk}}]{McDonald:05}
{McDonald}, P. {et~al.} 2006, \apjs, 163, 80,
  \adsurl{http://adsabs.harvard.edu/abs/2006ApJS..163...80M},
  \eprint{arXiv:astro-ph/0405013}

\bibitem[{{Mewe} {et~al.}(1986){Mewe}, {Lemen}, \& {van den Oord}}]{Mewe:86}
{Mewe}, R., {Lemen}, J.~R., \& {van den Oord}, G.~H.~J. 1986, \aaps, 65, 511,
  \adsurl{http://adsabs.harvard.edu/cgi-bin/nph-bib_query?bibcode=1986A
.65..511M&db_key=AST}

\bibitem[{{Miranda} \& {Macci{\`o}}(2007)}]{Miranda:07}
{Miranda}, M., \& {Macci{\`o}}, A.~V. 2007, 706,
  \adsurl{http://adsabs.harvard.edu/abs/2007arXiv0706.0896M},
  \eprint{0706.0896}

\bibitem[{{Moore} {et~al.}(1999){Moore}, {Quinn}, {Governato}, {Stadel}, \&
  {Lake}}]{Moore:99}
{Moore}, B., {Quinn}, T., {Governato}, F., {Stadel}, J., \& {Lake}, G. 1999,
  \mnras, 310, 1147,
  \adsurl{http://adsabs.harvard.edu/cgi-bin/nph-bib_query?bibcode=1999MNRAS.31%
0.1147M&db_key=AST}, \eprint{astro-ph/9903164}

\bibitem[{{Morrison} \& {McCammon}(1983)}]{Morrison:83}
{Morrison}, R., \& {McCammon}, D. 1983, \apj, 270, 119,
  \adsurl{http://adsabs.harvard.edu/abs/1983ApJ...270..119M}

\bibitem[{Navarro {et~al.}(1997)Navarro, Frenk, \& White}]{Navarro:96}
Navarro, J.~F., Frenk, C.~S., \& White, S. D.~M. 1997, \apj, 490, 493,
  \eprint{astro-ph/9611107}

\bibitem[{{Navarro} {et~al.}(2004){Navarro}, {Hayashi}, {Power}, {Jenkins},
  {Frenk}, {White}, {Springel}, {Stadel}, \& {Quinn}}]{Navarro:04}
{Navarro}, J.~F. {et~al.} 2004, \mnras, 349, 1039,
  \adsurl{http://adsabs.harvard.edu/abs/2004MNRAS.349.1039N},
  \eprint{arXiv:astro-ph/0311231}

\bibitem[{Nevalainen {et~al.}(2005)Nevalainen, Markevitch, \&
  Lumb}]{Nevalainen:05}
Nevalainen, J., Markevitch, M., \& Lumb, D. 2005, \apj, 629, 172,
  \eprint{astro-ph/0504362}

\bibitem[{Pal \& Wolfenstein(1982)}]{Pal:81}
Pal, P.~B., \& Wolfenstein, L. 1982, Phys. Rev., D25, 766

\bibitem[{Palazzo {et~al.}(2007)Palazzo, Cumberbatch, Slosar, \&
  Silk}]{Palazzo:07}
Palazzo, A., Cumberbatch, D., Slosar, A., \& Silk, J. 2007,
  \eprint{arXiv:0707.1495 [astro-ph]}

\bibitem[{{Peebles}(1980)}]{Peebles:80}
{Peebles}, P.~J.~E. 1980, {The large-scale structure of the universe}
  (Princeton, N.J., Princeton University Press, 1980.~435 p.),
  \adsurl{http://adsabs.harvard.edu/abs/1980lssu.book.....P}

\bibitem[{{Peebles}(1984)}]{Peebles:84a}
---. 1984, Science, 224, 1385,
  \adsurl{http://adsabs.harvard.edu/abs/1984Sci...224.1385P}

\bibitem[{{Penarrubia} {et~al.}(2007){Penarrubia}, {McConnachie}, \&
  {Navarro}}]{Penarrubia:07}
{Penarrubia}, J., {McConnachie}, A., \& {Navarro}, J.~F. 2007,
  \adsurl{http://adsabs.harvard.edu/abs/2007astro.ph..1780P},
  \eprint{astro-ph/0701780}

\bibitem[{Read \& Ponman(2003)}]{Read:03}
Read, A.~M., \& Ponman, T.~J. 2003, \aap, 409, 395,
  \adsurl{http://adsabs.harvard.edu/cgi-bin/nph-bib_query?bibcode=2003A
409..395R&db_key=AST}, \eprint{astro-ph/0304147}

\bibitem[{{Riemer-S{\o}rensen} {et~al.}(2006){Riemer-S{\o}rensen}, {Hansen}, \&
  {Pedersen}}]{Riemer:06}
{Riemer-S{\o}rensen}, S., {Hansen}, S.~H., \& {Pedersen}, K. 2006, \apjl, 644,
  L33,
  \adsurl{http://adsabs.harvard.edu/cgi-bin/nph-bib_query?bibcode=2006ApJ...64%
4L..33R&db_key=AST}, \eprint{astro-ph/0603661}

\bibitem[{Ruchayskiy(2007)}]{Ruchayskiy:07}
Ruchayskiy, O. 2007, in Proceedings of the 11th Marcel Grossmann Meeting on
  General Relativity, ed. H.~Kleinert, R.~Jantzen, \& R.~Ruffini (World
  Scientific), \eprint{arXiv:0704.3215 [astro-ph]}

\bibitem[{{S{\'a}nchez-Salcedo} {et~al.}(2006){S{\'a}nchez-Salcedo},
  {Reyes-Iturbide}, \& {Hernandez}}]{Sanchez:06b}
{S{\'a}nchez-Salcedo}, F.~J., {Reyes-Iturbide}, J., \& {Hernandez}, X. 2006,
  \mnras, 370, 1829,
  \adsurl{http://adsabs.harvard.edu/abs/2006MNRAS.370.1829S},
  \eprint{arXiv:astro-ph/0601490}

\bibitem[{Seljak {et~al.}(2006)Seljak, Makarov, McDonald, \& Trac}]{Seljak:06}
Seljak, U., Makarov, A., McDonald, P., \& Trac, H. 2006, Phys. Rev. Lett., 97,
  191303, \eprint{astro-ph/0602430}

\bibitem[{{Seto} \& {Yamaguchi}(2007)}]{Seto:07}
{Seto}, O., \& {Yamaguchi}, M. 2007, \prd, 75, 123506,
  \adsurl{http://adsabs.harvard.edu/abs/2007PhRvD..75l3506S},
  \eprint{arXiv:0704.0510}

\bibitem[{Shaposhnikov \& Tkachev(2006)}]{Shaposhnikov:06}
Shaposhnikov, M., \& Tkachev, I. 2006, Phys. Lett., B639, 414,
  \eprint{hep-ph/0604236}

\bibitem[{Shi \& Fuller(1999)}]{Shi:98}
Shi, X.-d., \& Fuller, G.~M. 1999, Phys. Rev. Lett., 82, 2832,
  \eprint{astro-ph/9810076}

\bibitem[{{Shirey} {et~al.}(2001){Shirey}, {Soria}, {Borozdin}, {Osborne},
  {Tiengo}, {Guainazzi}, {Hayter}, {La Palombara}, {Mason}, {Molendi},
  {Paerels}, {Pietsch}, {Priedhorsky}, {Read}, {Watson}, \& {West}}]{Shirey:01}
{Shirey}, R. {et~al.} 2001, \aap, 365, L195,
  \adsurl{http://adsabs.harvard.edu/abs/2001A\%26A...365L.195S},
  \eprint{arXiv:astro-ph/0011244}

\bibitem[{{Simon} \& {Geha}(2007)}]{Simon:07}
{Simon}, J.~D., \& {Geha}, M. 2007, 706,
  \adsurl{http://adsabs.harvard.edu/abs/2007arXiv0706.0516S},
  \eprint{0706.0516}

\bibitem[{{Smith} {et~al.}(2001){Smith}, {Brickhouse}, {Liedahl}, \&
  {Raymond}}]{Smith:01}
{Smith}, R.~K., {Brickhouse}, N.~S., {Liedahl}, D.~A., \& {Raymond}, J.~C.
  2001, \apjl, 556, L91,
  \adsurl{http://adsabs.harvard.edu/abs/2001ApJ...556L..91S},
  \eprint{arXiv:astro-ph/0106478}

\bibitem[{{Sommer-Larsen} \& {Dolgov}(2001)}]{Sommer:99}
{Sommer-Larsen}, J., \& {Dolgov}, A. 2001, \apj, 551, 608,
  \adsurl{http://adsabs.harvard.edu/abs/2001ApJ...551..608S},
  \eprint{arXiv:astro-ph/9912166}

\bibitem[{{Stanek} \& {Garnavich}(1998)}]{Stanek:98}
{Stanek}, K.~Z., \& {Garnavich}, P.~M. 1998, \apjl, 503, L131+,
  \adsurl{http://adsabs.harvard.edu/abs/1998ApJ...503L.131S},
  \eprint{arXiv:astro-ph/9802121}

\bibitem[{Stasielak {et~al.}(2007)Stasielak, Biermann, \&
  Kusenko}]{Stasielak:06}
Stasielak, J., Biermann, P.~L., \& Kusenko, A. 2007, \apj, 654, 290,
  \adsurl{http://adsabs.harvard.edu/abs/2007ApJ...654..290S},
  \eprint{arXiv:astro-ph/0606435}

\bibitem[{{Strigari} {et~al.}(2007){Strigari}, {Bullock}, {Kaplinghat},
  {Diemand}, {Kuhlen}, \& {Madau}}]{Strigari:07}
{Strigari}, L.~E., {Bullock}, J.~S., {Kaplinghat}, M., {Diemand}, J., {Kuhlen},
  M., \& {Madau}, P. 2007, 704,
  \adsurl{http://adsabs.harvard.edu/abs/2007arXiv0704.1817S},
  \eprint{0704.1817}

\bibitem[{{Str{\"u}der} {et~al.}(2001){Str{\"u}der}, {Briel}, {Dennerl},
  {Hartmann}, {Kendziorra}, {Meidinger}, {Pfeffermann}, {Reppin}, {Aschenbach},
  {Bornemann}, {Br{\"a}uninger}, {Burkert}, {Elender}, {Freyberg}, {Haberl},
  {Hartner}, {Heuschmann}, {Hippmann}, {Kastelic}, {Kemmer}, {Kettenring},
  {Kink}, {Krause}, {M{\"u}ller}, {Oppitz}, {Pietsch}, {Popp}, {Predehl},
  {Read}, {Stephan}, {St{\"o}tter}, {Tr{\"u}mper}, {Holl}, {Kemmer}, {Soltau},
  {St{\"o}tter}, {Weber}, {Weichert}, {von Zanthier}, {Carathanassis}, {Lutz},
  {Richter}, {Solc}, {B{\"o}ttcher}, {Kuster}, {Staubert}, {Abbey}, {Holland},
  {Turner}, {Balasini}, {Bignami}, {La Palombara}, {Villa}, {Buttler},
  {Gianini}, {Lain{\'e}}, {Lumb}, \& {Dhez}}]{Strueder:01}
{Str{\"u}der}, L. {et~al.} 2001, \aap, 365, L18

\bibitem[{Strumia \& Vissani(2006)}]{Strumia:06}
Strumia, A., \& Vissani, F. 2006, \eprint{hep-ph/0606054}

\bibitem[{{Takahashi} {et~al.}(2004){Takahashi}, {Okada}, {Kokubun}, \&
  {Makishima}}]{Takahashi:04}
{Takahashi}, H., {Okada}, Y., {Kokubun}, M., \& {Makishima}, K. 2004, \apj,
  615, 242, \adsurl{http://adsabs.harvard.edu/abs/2004ApJ...615..242T},
  \eprint{arXiv:astro-ph/0408305}

\bibitem[{Tegmark {et~al.}(2004)}]{Tegmark:03}
Tegmark, M., {et~al.} 2004, Phys. Rev., D69, 103501, \eprint{astro-ph/0310723}

\bibitem[{{Tempel} {et~al.}(2007){Tempel}, {Tamm}, \& {Tenjes}}]{Tempel:07}
{Tempel}, E., {Tamm}, A., \& {Tenjes}, P. 2007, 707,
  \adsurl{http://adsabs.harvard.edu/abs/2007arXiv0707.4374T},
  \eprint{0707.4374}

\bibitem[{Tremaine \& Gunn(1979)}]{Tremaine:79}
Tremaine, S., \& Gunn, J.~E. 1979, Phys. Rev. Lett., 42, 407

\bibitem[{{Turner} {et~al.}(2001){Turner}, {Abbey}, {Arnaud}, {Balasini},
  {Barbera}, {Belsole}, {Bennie}, {Bernard}, {Bignami}, {Boer}, {Briel},
  {Butler}, {Cara}, {Chabaud}, {Cole}, {Collura}, {Conte}, {Cros}, {Denby},
  {Dhez}, {Di Coco}, {Dowson}, {Ferrando}, {Ghizzardi}, {Gianotti}, {Goodall},
  {Gretton}, {Griffiths}, {Hainaut}, {Hochedez}, {Holland}, {Jourdain},
  {Kendziorra}, {Lagostina}, {Laine}, {La Palombara}, {Lortholary}, {Lumb},
  {Marty}, {Molendi}, {Pigot}, {Poindron}, {Pounds}, {Reeves}, {Reppin},
  {Rothenflug}, {Salvetat}, {Sauvageot}, {Schmitt}, {Sembay}, {Short},
  {Spragg}, {Stephen}, {Str{\"u}der}, {Tiengo}, {Trifoglio}, {Tr{\"u}mper},
  {Vercellone}, {Vigroux}, {Villa}, {Ward}, {Whitehead}, \&
  {Zonca}}]{Turner:01}
{Turner}, M.~J.~L. {et~al.} 2001, \aap, 365, L27,
  \eprint{arXiv:astro-ph/0011498}

\bibitem[{{Viel} {et~al.}(2007){Viel}, {Becker}, {Bolton}, {Haehnelt}, {Rauch},
  \& {Sargent}}]{Viel:07}
{Viel}, M., {Becker}, G.~D., {Bolton}, J.~S., {Haehnelt}, M.~G., {Rauch}, M.,
  \& {Sargent}, W.~L.~W. 2007, 709,
  \adsurl{http://adsabs.harvard.edu/abs/2007arXiv0709.0131V},
  \eprint{0709.0131}

\bibitem[{Viel {et~al.}(2006)Viel, Lesgourgues, Haehnelt, Matarrese, \&
  Riotto}]{Viel:06}
Viel, M., Lesgourgues, J., Haehnelt, M.~G., Matarrese, S., \& Riotto, A. 2006,
  Phys. Rev. Lett., 97, 071301, \eprint{astro-ph/0605706}

\bibitem[{Watson {et~al.}(2006)Watson, Beacom, Yuksel, \& Walker}]{Watson:06}
Watson, C.~R., Beacom, J.~F., Yuksel, H., \& Walker, T.~P. 2006, Phys. Rev.,
  D74, 033009, \eprint{astro-ph/0605424}

\bibitem[{{White} {et~al.}(1983){White}, {Frenk}, \& {Davis}}]{White:83}
{White}, S.~D.~M., {Frenk}, C.~S., \& {Davis}, M. 1983, \apjl, 274, L1,
  \adsurl{http://adsabs.harvard.edu/abs/1983ApJ...274L...1W}

\bibitem[Widrow \& Dubinski(2005)]{Widrow:05} Widrow, L.~M., \&
Dubinski, J.\ 2005, \apj, 631, 838,
\adsurl{http://adsabs.harvard.edu/abs/2005ApJ...631..838W},
\eprint{arXiv:astro-ph/0506177}

\bibitem[{{Wu}(2007)}]{Wu:07}
{Wu}, X. 2007,
  \adsurl{http://adsabs.harvard.edu/cgi-bin/nph-bib_query?bibcode=2007astro.ph%
..2233W&db_key=PRE}, \eprint{astro-ph/0702233}

\bibitem[{{Zel'dovich}(1970)}]{Zeldovich:70}
{Zel'dovich}, Y.~B. 1970, \aap, 5, 84,
  \adsurl{http://adsabs.harvard.edu/abs/1970A\%26A.....5...84Z}

\end{thebibliography}
\end{document}